\documentstyle[12pt,preprint]{aastex}
\begin{document}

\title{The Outer Disk of the Milky Way Seen in $\lambda$21-cm Absorption}

\author{John M. Dickey}
\affil{University of Tasmania}
\affil{School of Maths and Physics, Private Bag 37, Hobart, TAS 7001, Australia}
\email{john.dickey@utas.edu.au}
\author{Simon Strasser}
\affil{University of Minnesota and Dominion Radio Astrophysical Observatory}
\author{B.M. Gaensler}
\affil{Institute of Astronomy, School of Physics, The University of Sydney}
\author{Marijke Haverkorn}
\affil{NRAO Jansky Fellow, University of California at Berkeley and Astron - Netherlands Institute for Radio Astronomy}
\author{Dain Kavars}
\affil{University of Minnesota and Ball State University}
\author{N. M. McClure-Griffiths}
\affil{Australia Telescope National Facility, CSIRO}
\author{Jeroen Stil, A. R. Taylor}
\affil{University of Calgary}

\keywords{ISM: atoms, ISM: clouds, ISM: structure, Galaxy: disk, Galaxy: structure}

\begin{abstract}
Three recent surveys of 21-cm line emission in the Galactic plane,
combining single dish and interferometer observations to achieve
resolution of 1\arcmin \ to 2\arcmin, $\sim 1$ km s$^{-1}$,  and good
brightness sensitivity, have provided some 650 absorption spectra with corresponding
emission spectra for study of the distribution of warm and cool phase
H{\tt I} in the interstellar medium.  These emission-absorption
spectrum pairs are used to study the temperature of the interstellar
neutral hydrogen in the outer disk of the Milky Way, outside the  
solar circle, to a radius of 25 kpc. 

The cool neutral medium is distributed in radius and height above
the plane with very similar parameters to the warm neutral medium.
In particular, the ratio of the emission to the absorption, which
gives the mean spin temperature of the gas, stays nearly constant with 
radius to $\sim$25 kpc radius.  This suggests that the mixture
of cool and warm phases is a robust quantity, and that the changes in
the interstellar environment do not force the H{\tt I} into a regime
where there is only one temperature allowed.  The mixture of atomic
gas phases in the outer disk is roughly 15\% to 20\% cool (40 K to 60 K),
the rest warm, corresponding to mean spin temperature $\sim 250$ to 400 K.

The Galactic warp appears clearly in the absorption data, and
other features on the familiar longitude-velocity diagram 
have analogs in absorption with even higher contrast than for
21-cm emission.  In the third and fourth Galactic quadrants the
plane is quite flat, in absorption as in emission, in contrast to
the strong warp in the first and second quadrants.  The scale
height of the cool gas is similar to that of the warm gas,
and both increase with Galactic radius in the outer disk.
\end{abstract}

\section{Background}

Surveys of the Milky Way disk using the 21-cm line have been one of the most powerful
means of tracing the structure and properties of the Galaxy for over 50 years \citep[reviewed
by][]{Burton_1988,Burton_1991,Lockman_2002,Kalberla_Dedes_2008,Kalberla_Kerp_2009}.  Our
knowledge of the outer Galaxy, beyond the solar circle,
is particularly dependent on H{\tt I} emission surveys, as emission from other species in the interstellar
medium (ISM) declines faster with Galactocentric radius, $R_g$.  Thus the 21-cm line is much easier to detect and to
use as an ISM tracer in the outer Galaxy than lines from molecules like CO, both because the
atomic phase of the medium is becoming more and more the dominant form of the gas mass with
increasing Galactic radius, $R_g$, and because the H{\tt I} has a higher spatial filling factor than the molecules.  

Since the pioneering H{\tt I} surveys of the 1950's it has been clear that the
Milky Way disk extends to at least two to three times the radius of the solar circle, $R_o$.
Surveys covering wide latitude ranges \citep{Burke_1957, Oort_Kerr_Westerhout_1958} showed
that the H{\tt I} disk in the outer Galaxy is warped: in the longitude range $50\arcdeg$ to $130\arcdeg$ 
the middle or centroid of the gas distribution
moves toward positive $z$, where $z$ is the height above the plane defined by latitude $b=0\arcdeg$.
In the third and fourth quadrants (longitudes $240\arcdeg$ to $310\arcdeg$)
there is very little displacement of the gas from this flat plane, at least
to $R_g\simeq 25$ kpc.   This does not necessarily imply an asymmetry in the Milky Way disk,
since our location places us much nearer the warp
in the first and second quadrants than its reflection on the other side of the Galactic
center, which would be expected at high longitudes in the fourth quadrant
\citep{Kalberla_etal_2007,Levine_Blitz_Heiles_2006}. 
At the same radii where the warp
becomes significant, $R_g \sim 15$ kpc, the gas disk begins to flare, meaning that its scale height increases.
This happens at all galactocentric azimuths, $\phi$, defined as zero in the direction of
longitude zero, i.e. a ray pointing from the Galactic center directly away from the sun.

Although surveys of H{\tt I} {\bf emission} at low Galactic latitudes have been done for many years and
with many telescopes,
surveys of {\bf absorption} in the 21-cm line have been much less common, because
the instrumental requirements to measure absorption are more stringent than for
emission, as discussed in section 2 below.  Interferometer and aperture synthesis telescopes
like the NRAO Very Large Array (VLA)
were used for low latitude absorption surveys in the 1970's and 80's \citep{Radhakrishnan_etal_1972,
Goss_etal_1972,Mebold_etal_1982, Dickey_etal_1983, Kolpak_etal_2002}, but interferometers
have limited capability to measure the 21-cm emission, so surveys of emission and absorption in
these decades were done
separately using single dish telescopes to measure the emission.  This necessarily gave very
different effective beam sizes for the emission and absorption spectra, which is problematic
for analysis that involves combining the two.

In the late 1990's and early 2000's three large surveys of the 21-cm emission at low
latitudes were undertaken: 
the Canadian Galactic Plane Survey \citep[CGPS,][]{Taylor_etal_2003}, the Southern Galactic
Plane Survey \citep[SGPS,][]{ McClure-Griffiths_etal_2005}, and the VLA Galactic Plane Survey
\citep[VGPS,][]{Stil_etal_2006}.  These surveys were the first to combine data from single dish and
aperture synthesis telescopes, for the latter applying the recently perfected 
mosaicing technique \citep{Sault_Staveley-Smith_Brouw_1996} for
recovering the short-spacing information.  This allowed
maps of the H{\tt I} to be made over wide areas with sensitivity to all spatial scales,
from many degrees down to the survey resolution of 45\arcsec (VGPS), 1\arcmin \ (CGPS)
or 2\arcmin \ (SGPS).  Thus the survey data is equivalent to a fully sampled map made with a single
dish telescope with this beamwidth; such a single dish would have to be 500m or more in diameter!
This resolution allows reasonably accurate measurement of the absorption spectra toward a large
number of continuum background sources at low latitudes, with corresponding emission interpolated
from spectra nearby taken with the same resolution.  The resulting emission-absorption
spectrum pairs are ideally suited for measurement of the excitation of the 21-cm line 
from interstellar gas throughout the Galactic plane.  This paper presents these
emission-absorption spectra from the three surveys, and discusses briefly
their implications for the thermodynamics of the hydrogen in the outer disk of the Milky Way.
A preceeding paper in this series \citep{Strasser_etal_2007} presents a preliminary
study of the absorption data from these surveys, and discusses the morphology
of the cool gas complexes seen at large $R_g$.

The main motivation to measure emission and absorption spectra with the same resolution in the
same directions is in order to determine the excitation temperature of the 21-cm line, called
the spin temperature, $T_{sp}=\frac{T_{EM}}{1-e^{-\tau}}$ where $T_{EM}(v)$ is the brightness
temperature of the H{\tt I} line as a function of radial velocity in the emission spectrum
and $\tau(v)$ is the optical depth of the H{\tt I} line.  In the Galactic environment $T_{sp}$
is generally close to 
the kinetic temperature \citep[see][for a review of the astrophysics of the spin temperature
in various environments]{Furlanetto_Oh_Briggs_2006}. 
Blending of different regions
with different temperatures along the line of sight can make the interpretation complicated, 
as discussed briefly in section 4 below, and more fully by \citet{Dickey_etal_2003}.  
Because of this blending of warm and cool gas at the same velocity, the 
measured value of $T_{sp}$ is generally higher than the temperature of
the cool clouds that are responsible for most of the absorption, $T_{cool}$.  
This bias is independent of distance.
Surveys of emission and absorption at low latitudes make it possible to map the spin temperature
throughout the Galaxy, using radial velocity as a kinematic distance indicator.  In the
outer Galaxy this is of particular interest, since the physical processes that dominate
heating and cooling in the atomic phase may change drastically with $R_g$.  Going from the
solar circle to $R_g\simeq 20$ kpc the density of stars in the disk drops by more than an order of magnitude,
and the acceleration due the gravity of the disk, $K_z$, drops similarly, which is 
the cause of the flaring or thickening of the gas layer.  Although the ISM
is not necessarily in hydrodynamic equilibrium with the gravitational potential of the 
disk on small scales, at least on long time scales the gas pressure cannot
be very different from that set by the gravitational force on the gas above \citep{Spitzer_1956}.
Thus the average pressure at mid-plane must drop by more
than an order of magnitude in the outer Galaxy compared to its solar circle value.
Depending on the metallicity gradient, the standard theory of H{\tt I} thermodynamic equilibrium
\citep{Wolfire_etal_1995, Wolfire_etal_2003} could predict that this pressure drop
would lead to an overall phase change, with
all the cool neutral medium (CNM) converting to warm neutral medium (WNM) at
some $R_g$.  A major result of this study, described in sections 3 and 4, 
is therefore something of a surprise, as we find that the mixture of CNM and WNM is robust, with
little or no change in the relative fractions of these two phases
with $R_g$ out to nearly three times $R_o$ or 25 kpc.

\section{Survey Data Reduction}

To measure absorption requires a continuum background source, and a way of
subtracting the emission of the gas toward the source so that the optical depth can be determined.
Since emission and absorption are mixed in the spectrum toward the background
source, an interpolation of the surrounding emission spectra must be
used to estimate and subtract the emission at the position of the continuum.  
The emission at the position of the continuum source is estimated in the
following way: 1) The structure of the continuum source is determined from the
line free channels. 2) The absorption is determined on source and 3) the
expected emission estimated from the close surroundings of the source.
Accurate estimates for emission requires a combination of single dish and
interferometer data.
At low Galactic latitudes the H{\tt I} emission shows random spatial variations on all angular
scales, so a small telescope beam is required in order for this interpolation to be made over
angles of a few arc minutes or less.  Typically the mean square emission fluctuations on a given
angle, $\theta$, are proportional to $\theta^{-3}$ to $\theta^{-4}$
\citep{Dickey_etal_2001}, so the error in the interpolated emission profile depends on the
beamwidth to a power between 1.5 and 2. For a given level of error in this interpolated 
emission spectrum, the resulting error in the absorption (1 - e$^{-\tau}$) goes
inversely as the continuum antenna temperature due to the background source.
This continuum antenna temperature is given by the source flux density times the
telescope gain ($G$ in K/Jy), which itself is inversely proportional to the square of the beam
width.  Thus the error in the measured absorption introduced by emission fluctuations
decreases with decreasing beam width to a power between three and four.
This effect dominates other sources of error such as ordinary radiometer noise
in most 21-cm absorption surveys, particularly at low latitudes.  
For example, the Arecibo telescope, with beamwidth of 3.2\arcmin, is able to measure
absorption spectra toward continuum sources of a few Jy or stronger at high and
intermediate latitudes \citep{Heiles_Troland_2003a}, but for latitudes
below about $10\arcdeg$ even this beam size is too large to give a sufficiently
accurate interpolated emission profile.

Absorption spectra may suffer more from emission fluctuations than the emission
spectra do, because an error in the interpolated emission that is subtracted
from the spectrum toward the continuum source can cause a large fractional error in
the optical depth.
Aperture synthesis telescopes can solve the problem of fluctuations in the emission
by performing a high-pass spatial filtering of the brightness distribution as set
by the {\it uv} plane sampling function of the telescope baselines.  Thus telescopes
like the VLA and the Australia Telescope Compact Array (ATCA)
can measure absorption toward compact continuum sources brighter than a few tens 
of mJy, because the longer baselines of these telescopes are sensitive only to structure
smaller than a few arc seconds.  This spatial filter can also reduce the continuum emission of
the background source, depending on its angular size, so except for the most
compact continuum sources there is a point of diminishing returns in the spatial
filtering by using longer and longer baselines to measure 21-cm absorption.
The SGPS data have been spatially filtered to improve the accuracy
of the absorption spectrum (but not for the emission spectrum, of course), 
the radius of the spatial filter is adjusted to leave the most continuum
while minimizing the spectral line emission that leaks through the spatial filter
due to small angle emission variations.  The effect of
these emission fluctuations in the absorption spectra is obvious because
they generate spurious ``absorption'' lines with a symmetric distribution
around zero in $\left(1-e^{-\tau}\right)$, i.e. equal numbers of spurious negative and positive
peaks in the optical depth.

Details of the techniques for doing the interpolation of the emission and
so measuring the emission and absorption are given by \citet{Strasser_Taylor_2004,
Dickey_etal_2003, McClure-Griffiths_etal_2001, Strasser_2006}.  These
techniques are slightly
different for the different surveys.  For the SGPS and VGPS surveys the
continuum was measured together with the spectral line channels, but for
the CGPS the continuum was measured separately.  
So for the CGPS absorption spectra the absolute calibration of the
optical depth has some errors in overall scale factor,
typically this is less than 5\% but in a few cases as high as 10\%.
Table 1 gives observational parameters of the different surveys.  
Note that the survey data provided by the web servers
on Table 1 have had the continuum
subtracted, whereas the analysis described here was necessarily done 
with data from an earlier stage of
the reduction, before the continuum subtraction.

The spectra are grouped according to the continuum antenna temperature
of the background source, which sets the noise in $\left(1-e^{-\tau}\right)$, as 
measured at frequencies away from the Galactic velocities where
there is no emission or absorption in the 21-cm line.  For
the brightest continuum sources, the noise in $\left(1-e^{-\tau}\right)$ is 
$\sigma_{\tau} \sim 10^{-2}$.  There are 77 spectrum pairs with
$\sigma_{\tau}<0.02$ in the three surveys combined.  Numbers with
$\sigma_{\tau}$ below 0.05 and 0.10 are given on table 1.  Although
the spectra in the third group (0.05$< \sigma_{\tau}<$0.1) are
not of high enough quality to measure accurate values for $T_{sp}$
in each velocity channel, they can give useful results for the 
integral of $\left(1-e^{-\tau}\right)$ over broader ranges of velocity, particularly
when averaged together with others in a sample of many spectra.
The critical point is that both for radiometer noise and for
errors in the absorption spectra due to fluctuations in the
emission, there is no bias toward positive or negative values of
$\left(1-e^{-\tau}\right)$.  


The emission and absorption spectrum pairs are shown in figures 1-3, ordered
by longitude, for the three groups defined by $\sigma_{\tau} \leq 0.02$,
$0.02 < \sigma_{\tau}\leq 0.05$, and $0.05 < \sigma_{\tau}\leq 0.10$.
The print edition shows just the first and last pair for
each survey in each group.  The left hand panel shows the
emission and absorption spectra, with LSR velocity
scaled across the bottom, and absorption $\left(1-e^{-\tau}\right)$ scaled
on the left hand axis.  To provide as much detail as possible
in the absorption profile, the velocity scale is expanded
to cover only the range where the emission is non-zero,
which necessarily includes all the absorption.  

The emission spectra (plotted in grey, or in gold on the
electronic edition) are offset upward from the absorption
spectra for clarity, and scaled in K of brightness temperature
on the right hand axis of the left hand panel, which is the
same scale as the vertical axes of the right hand panel.
The bottom axis of the right hand panel is again $\left(1-e^{-\tau}\right)$.
Thus the right hand panel shows emission plotted against 
absorption.  These plots are crowded in many cases, but the
points on the right panel can be identified with velocity
channels on the left panel by drawing a horizontal line
through the $T_{EM}$ axis between the two panels.
Analysis of the separate loops on the right hand panel
allows the temperature of the cool gas traced by the
absorption to be determined, the remaining emission is
due to warm gas blended in velocity with the cool gas,
some of which may be associated with the 
cool clouds and some not \citep{Mebold_etal_1997,Dickey_etal_2003}.
This analysis is beyond the scope of this paper, but it
has been performed by \citet{Strasser_2006} for all the 
absorption features in all the spectra.

%

\section{Results}
\subsection{Longitude - Velocity Diagrams}

The absorption spectra are necessarily taken where the
background sources are, which does not give a regular and
fully sampled map of the optical depth.  In order to
study the Galactic distribution of the cool gas that
causes the absorption, we bin the spectra into grids
of longitude and velocity as shown on figure 4.
These show pairs of longitude-velocity diagrams, using
the emission-absorption spectrum pairs, with all spectra
contributing to each bin averaged together.  The longitude
step size is 1.8 degrees and the velocity step size
is 2 km s$^{-1}$.  Typically, more than one spectrum contributes
to each bin.  Spectra from the full latitude range of each survey 
are included.

The longitude-velocity diagrams of the emission (figure 4)
are similar to plots from emission surveys
\citep[e.g.][]{McClure-Griffiths_etal_2005} which allow structures in
the density and velocity distributions of the ISM to be
studied in many ways \citep[reviewed by][]{Burton_1988}.  
In this case the relatively sparse sampling provided by
the background continuum source directions makes the 
{\it l}-v diagram appear crude by modern standards, but the
major features are clear.  Figure 4
shows two versions of each {\it l}-v diagram, with two different
grey scales, to make these visible.  On the upper panel,
curves of constant galactocentric distance, $R_g$, are
indicated, based on a flat rotation curve outside the
solar circle.  The flat rotation curve approximation gives
simple equations for the kinematic distance, $d(v)$, and
hence for $R_g$ and $\phi$ from the laws of cosines and
sines, respectively:

\begin{equation} \frac{d}{R_0} \ = \ \cos{\lambda} \ + \ \sin{\lambda} \ \sqrt{
\left( \frac{v}{\cos{b}} \ + \ \sin{l} \right)^{-2} \ - \ 1} \end{equation}

\begin{equation} \frac{R_g}{R_0} \ = \ \sqrt{ 1 \ + \ d^2 \ - \ 2d \cos{\lambda}} \end{equation}

\begin{equation} \phi \ = \ \arcsin{\left( \sin{\lambda} \ \frac{d}{R_g} \right)} \end{equation}

\noindent
where $d(v)$ is the distance (from the sun) as a function of the
observed radial velocity, $v$,
in units of $v_o \equiv 220$ km s$^{-1}$, the LSR velocity around the solar
circle (which is assumed to be the
circular velocity for all $R_g > R_0$), and $\lambda$ is equal to the
longitude, $l$, in the first and second quadrants
but $\lambda\ = \ -l$ in the third and fourth
quadrants.  Here $d$ and $R$ 
are both in units of $R_o \equiv 8.5$ kpc. 
Note that $R_o$ and $v_o$ are simply scale
factors that do not otherwise effect the analysis, and
even the choice of a flat rotation curve is not
much more significant than a choice of scale.  Using any
other smooth monotonic function for the circular rotation
velocity vs. R would cause a smooth stretching
of the scale of R, but otherwise the results of
the analysis would be unchanged.  The upper right panel
of figure 4
shows for comparison distances derived from a rotation
curve that is flat to 15 kpc and then drops
as a Keplerian.  This certainly under-estimates
the distance to gas at a given radial velocity,
but the displacement of the curves is less than
20 km s$^{-1}$ compared with those on the top left.

The lower panels of figure 4 show, with different
grey scales, the longitude-velocity diagram of $(1-e^{-\tau})$,
the optical depth of the 21-cm line.  The structures seen
in the absorption correspond to structure seen in emission,
but with more contrast between arm and interarm regions. 
This is clear in spite of the higher noise level in the optical depth.
This may be partly due to the fact that the linewidths
seen in absorption are narrower than those of H{\tt I} 
emission lines; typically the full width at half maximum
of an absorption line is ~4 km s$^{-1}$ whereas emission features
have widths of 10 to 20 km s$^{-1}$.  In addition, the volume
filling factor of the CNM that causes the H{\tt I} absorption
is much less than that of the WNM seen in emission, roughly
3 to 5\% vs. roughly 50\% \citep{Kalberla_Kerp_2009}.  So for many reasons the 
H{\tt I} absorption traces better than the emission the structures
seen in cold gas, on large scales as seen on the {\it l}-v diagram
as well as on smaller scales in individual spectrum pairs.
In this way the 21-cm absorption has some of the characteristics
of molecular line tracers such as $^{12}$CO and $^{13}$CO emission
\citep{Jackson_etal_2002}.

The {\it l}-v diagrams of the emission and absorption are
shown in a different format by \citet[figure 3]{Strasser_etal_2007}.
Comparison between that technique, which simply plots the absorption
spectra at their respective longitudes with color-coding for
the depth of the absorption, with the binning technique used
in figure 4 is interesting.  In \citet{Strasser_etal_2007}
the corresponding {\it l}-v diagram of the emission includes
all data, not simply the directions toward the background
sources used for the absorption, so it has much finer 
resolution in longitude.  As discussed in that paper, 
absorption surveys provide vital information for tracing
large scale features in the outer disk, such as spiral arms.

\subsection{Radial Distribution of the Opacity}

To study the distribution of the CNM on the largest scales 
in the outer Galaxy, we transform the velocity into distance
from the Galactic center using equations 1 - 3, and
then average over annuli of constant $R_g$.  This gives 
figure 5.  On each figure the data from the three
surveys is plotted separately.  This separates data from the lower
longitudes in the first quadrant, $18\arcdeg  \leq  l  \leq 65\arcdeg$  
in the VGPS from the higher longitudes in the first and second 
quadrant, $65\arcdeg  \leq  l  \leq  170\arcdeg$ in the CGPS, and
the third and fourth quadrant, $255\arcdeg  \leq  l  \leq  357\arcdeg$
of the SGPS.  These are plotted in gold, black, and red respectively
on the electronic edition.  The x axis of figure 5 shows
$R_g$ ranging from 8.5 to 25 kpc in steps of 0.1 $R_o$ or 0.85 kpc. 
The y axis on the top panel of figure 5 plots the
log (base 10) of the average of the emission 
brightness temperature (in K), times the spectral channel band width,
$\Delta v$ in km s$^{-1}$, divided by the
line of sight path length, $\Delta L$ in kpc, corresponding to
that range of velocity:

\begin{equation} \left< \frac{T_{EM} \ \Delta v}{L} \right> \ = \ \frac{\sum_i T_{i} \ 
\frac{\Delta v}{\Delta L} \  w_i}{\sum_i w_i} \end{equation}

\noindent
where the sum is taken over many spectra, and many
spectral channels, $i$, in each spectrum, that correspond
to the given range of $R_g$, and 
$w_i$ is a weight factor set by the 
inverse of the noise in each absorption spectrum.
The path length corresponding to the velocity width of one spectral channel is
$\Delta L = \left| d(v + \Delta v) - d(v)\right|$ 
at the longitude and velocity of channel $i$.
We experimented with different choices for noise weighting, 
$w_i$, converging on the inverse of the rms noise in
$\left(1-e^{-\tau}\right)$ as the most robust.  Of course this
weighting is designed to optimize the signal to noise of
the resulting average only for the {\bf optical depth}, shown on the 
middle panel of figure 5, but we use the same 
weighting for the emission so that the two averages can be
directly compared with all other parameters kept the same.

The brightness temperature average in equation 4 has units
K km s$^{-1}$ kpc$^{-1}$, but this can be converted to density of  
H{\tt I} since the brightness temperature averaged over a velocity step
traces the column density of gas (for optically thin 21-cm emission):

\begin{equation} \left< \frac{n_H}{{\rm cm}^{-3}} \right> \ =
\ \frac{N_H}{\Delta L} \ = \ 
\ 5.9 \ 10^{-4} \ \left<
\frac{T_{EM} \ \Delta v}{{\rm K\ km\ s}^{-1}}
\ \left(\frac{L}{\rm kpc}\right)^{-1} \right>
\end{equation}

\noindent
with $\Delta L$ the path length interval, $N_H$ the column density
of H{\tt I} and $n_H$ the average space density
along this path.  
Thus the value 3 on the y axis of the top panel on figure 5, meaning
$10^3$ K km s$^{-1}$ kpc$^{-1}$, corresponds to
density 0.59 cm$^{-3}$, which is roughly the midplane density at the
solar circle.  The density drops rapidly
outside the solar circle, with exponential scale length about 3.1 kpc.
This is generally consistent with the much better determined
values for these numbers from large scale H{\tt I} emission surveys
\citep{Kalberla_Dedes_2008}.  The strong departures from a smooth
radial decrease in the CGPS and VGPS data on figure 5 are due
to a combination of the Perseus arm, which increases the H{\tt I} density
above its mean value at $R_g \simeq$ 10 to 15 kpc in
the first and second quadrants, and the warp,
that strongly reduces the H{\tt I} density in the VGPS longitude range at 
$R_g > 15$ kpc.  The SGPS data show weaker departures from
the underlying exponential decrease, because the effects of
the warp are much less significant in the third and fourth quadrants.
  
The middle panel of figure 5 shows the radial
variation of the opacity in the H{\tt I} line,
i.e. the absorption equivalent width per unit line of sight length,
$\left< \kappa \right>$,
which is the major new result of this study.  The y axis now plots
the average of the absorption, $\left(1-e^{-\tau}\right)$, per spectral channel,
weighted in exactly the same way as for the emission in equation 4, giving
the average of the ratio of the density divided by the spin temperature:

\begin{equation} \left< \frac{n_H}{T_{sp}}\right> \  \left( \frac{K}{{\rm cm}^{-3}}
\right) \ = \ \left< \frac{N_H}{T_{sp}} \frac{1}{L} \right> \left( \frac{K cm}{cm^{-2}}
\right) \ = \ \ 5.9 \ 10^{-4} \ \left< \frac{\tau \ \Delta v}{{\rm  km\ s}^{-1}}
\ \left(\frac{L}{\rm kpc}\right)^{-1} \right>
\end{equation}

\noindent
with the quantities as in equation 5, and $T_{sp}$ the
excitation temperature of the 21-cm transition.  
The term in brackets in the right
hand expression in equation 6 is the opacity, $\left< \kappa \right>$,
where the brackets denote a line of sight average over all phases of the ISM. 
In this analysis we work with
the directly observed quantity $\left(1-e^{-\tau}\right)$ rather than $\tau$,
so that the noise is not amplified at velocities where the optical
depth is significant, this substitution 
causes the values to be slight underestimates.
This is not
significant in the outer Galaxy where the optical depth is generally
much less than one.  On the y axis of the middle panel of
figure 5, the value 1, meaning $\left< \kappa \right> = 
10^1$ km s$^{-1}$ kpc$^{-1}$, now corresponds to,
e.g., density 0.59 cm$^{-3}$ and spin temperature 100 K.  The average
of $T_{sp}$ over WNM and CNM at the 
solar circle is between 150 and 250 K \citep{Dickey_etal_2000}, so the
expected value for $\frac{n}{T_{sp}}$ is
about 3 10$^{-3}$ cm$^{-3}$ K$^{-1}$, 
corresponding to 0.7 on the left-hand axis of figure 5, middle panel.
Absorption surveys of the inner Galaxy typically give numbers
of 5 to 10 km s$^{-1}$ kpc$^{-1}$ for $\left< \kappa  \right>$, which is
consistent with the data from the three surveys at the left edge
of the middle panel of figure 5
\citep{Kolpak_etal_2002,Garwood_Dickey_1989}.

The radial distributions of the emission and the absorption can
be combined into an effective spin temperature by dividing the
azimuthal averages of the emission brightness temperature per unit
line of sight distance, shown on the top panel of figure 5, by the optical depth
per unit distance shown on the middle panel.  The result is shown on the
bottom panel.
The y axis now is the average excitation temperature, computed not
by dividing the emission by the absorption channel-by-channel, but
by dividing the averages of many channels from many spectra that fall
in the same radial bins.  The result is surprisingly constant with
Galactic radius.  Whereas the emission and absorption alone 
decrease by some two orders of magnitude over the radial range
10 to 25 kpc, including departures from the smooth exponential 
by at least a factor of three at certain longitudes, yet their
ratio stays constant within a factor of two over this entire radial
range, with the exception of the VGPS at $R_g > 17$ kpc, for which
the optical depths are so small, due to the warp displacing the
plane out of the survey area, that the absorption is not detected
above the noise in this survey at these radii.

On the bottom panel of figure 5 the SGPS and VGPS
show very good agreement with $\left<T_{sp}\right>
\simeq 400$ K for $10 \leq R_g \leq 17$ kpc.  The CGPS shows significantly
cooler values of $\left<T_{sp}\right> \simeq 250$ K over the same range.  It is 
possible that this is due to the geometry of the warp, or some other
factor that makes the CNM more abundant in the CGPS area than in the
areas covered by the other surveys.  It may also reflect in part a
bias in the continuum estimate used to compute the optical depth in the
CGPS survey, that could arise from the separate processing of the
spectral line and continuum images in that survey.  For this
reason, the absolute
calibration of the absorption spectra from the VGPS and SGPS is
probably more reliable than that of the CGPS, even though there are
many more lines of sight sampled in the CGPS, which leads to smaller
error bars on the lower two panels of figure 5.

In all three panels of figure 5 the error bars represent the rms dispersion 
of the results for the
three groups of spectra, those with $\sigma_{\tau} < 0.02$, those
with $0.02 \leq \sigma_{\tau} < 0.05$, and those with
$0.05 \leq \sigma_{\tau} < 0.1$, averaged without weighting
by $\sigma_{\tau}$.  The small symbols mark
the weighted average of all spectra from all three groups.

\subsection{Variation of the Opacity with Azimuth, {\bf $\phi$}}

The radial distribution of the H{\tt I} 
opacity, $\left< \kappa \right>$, shows broad general agreement among
the surveys, but the effect of the warp is clearly quite significant
in the VGPS longitude range, and in the lower longitudes of the CGPS 
as well.  To see this better, we bin the data in radius and azimuth,
as shown in figures 6a and 6b.  Here the disk is plotted as seen 
face-on from above the North Galactic Pole.  The blank circle at the centre
has radius $R_0$, and the maximum radius plotted is $R_g$ = 25 kpc.
These figures show in grey-scale the
same quantities plotted on the y-axis of the upper
two panels of figures 5, i.e. the
log of the 21-cm emission per unit path length, proportional to $n_H$,
and the log of the 21-cm absorption per unit path length, proportional
to $\frac{n_H}{T_{sp}}$.  The grey scale extends over a range of 
$\sim 10^3$ in each quantity, slightly less than the full scale on the
y-axes of the upper two panels on figure 5.  The effect of the warp in the first
and second quadrants is seen as the white boxes in the lower right
corners of each figure.

The azimuthal bins on figures 6a and 6b are ten degrees in angle by
0.1 times $R_0$ in radius.
Even with this relatively coarse sampling, the number of lines
of sight passing through some bins is not very large, so there is
quite a bit of fluctuation from bin to bin, due to small number
statistics.  The overall radial trend shown on figure 5 
is clearly evident, as is the Perseus Arm in the first and second
quadrant, which doubles both the H{\tt I} density and opacity.  In the
third and fourth quadrant there is weak indication of the distant
arm mapped in 21-cm emission by \citet{McClure-Griffiths_etal_2004}.

\subsection{Variation with Height Above the Plane, {\bf $z$}}

Figures 7a and 7b show the dependance of the H{\tt I} density and 
opacity on $z$, the height above or below the plane defined by
b=0\arcdeg.  In order to have a sufficiently large number of measurements
contributing to each 100 pc wide interval in $z$, the data in
these figures are averaged over ranges in azimuth and
radius.  Figure 7a shows a histogram of the distribution 
of the density of H{\tt I} from equations 4 and 5, as a function
of $z$ over the range -1.1$< z <$+1.1 kpc.
The different panels show averages over separate regions
of $R_g$ and $\phi$, with the top row using only data with
$0\arcdeg \leq \phi < 100\arcdeg$, the middle row using only data with
$100\arcdeg \leq \phi < 180$\arcdeg, and the bottom row using only data with
$180\arcdeg \leq \phi < 360$\arcdeg.  The data is also separated into
four radial ranges, corresponding to the four columns.
The left column uses data from the range $1.0 < \frac{R_g}{R_0} \leq 1.2$
the next uses data from the range $1.2 < \frac{R_g}{R_0} \leq 1.4$
the next uses data from the range $1.4 < \frac{R_g}{R_0} \leq 2.0$
and the right column uses data from the range $2.0 < \frac{R_g}{R_0} \leq 3.0$.
Figure 7b shows the distribution of $\left< \kappa \right>$, the mean
opacity in equation 6, for the same ranges of $\phi$ and $R_g$.  

The histograms on figures 7a and 7b indicate the relative amounts
of gas at each height, $z$, in each region.  They are all normalized
so that the largest bar has value 1.0.  The empty boxes, drawn
in red in the on-line edition, indicate that the number of 
measurements in that bin is zero, or too small to give a 
reliable average.  Note that the absolute scale is such
that the sum of the histograms in each column would match the corresponding
radial averages on figures 5, thus the scale expands
by a factor of $\sim$30 going from the left column to the right column.
The rightmost column shows the effects of noise in the optical
depth, as there are several negative histogram bars as well
as positive ones.

The shapes of the histograms in corresponding panels of figures 7a and 7b show
a strong similarity, meaning that
the distributions of the emission, proportional to $n_H$, and of the
opacity, proportional to $\frac{n_H}{T_{sp}}$, are very
similar.  This is surprising, since at the solar circle and
in the inner Galaxy the scale height of the CNM is smaller
than that of the WNM \citep{Dickey_Lockman_1990, McClure-Griffiths_Dickey_2007}.  The
implication is that the flaring or thickening of the disk in $z$ as 
$R_g$ increases in the outer galaxy, that makes the widths
of the distributions increase going from left to right on
figures 7a and 7b, is a process that effects the WNM and
CNM with equal strength.  As \citet{Liszt_1983} shows, some
of the warm phase gas must be associated with the cool gas,
while some is in a more widely spread, neutral intercloud medium.
The fact that the scale height of the absorption is only
slightly smaller than that of the emission in the outer
Galaxy suggests that the scale heights of these two warm H{\tt I}
components are not very different.

The effect of the warp is seen strongly in the top row
of panels in both figures 7a and 7b.  The first moment
or mean value of $z$ increases with $R_g$, finally moving
out of the survey window altogether for $R_g > 15$ kpc.
This is the same effect that causes the empty bins in the
lower right area of figures 6a and 6b.  In the SGPS data, 
shown in the bottom row of panels on figures 7a and
7b, there is very little departure from the midplane in
the averages over $180\arcdeg < \phi < 360$\arcdeg.

\section{Conclusions}

The constancy of the mean spin temperature with Galactic radius
from $R_0$ to 3$\times R_0$, as shown on the bottom panel of figure 5, indicates
that the mixture of CNM and WNM is roughly the same throughout
the outer Milky Way disk.  This is surprising, given that the
physical conditions, in particular the gas pressure, the supernova rate, the 
radiation field, and the heavy element abundance, are all 
decreasing with $R_g$, and these variations must influence
the balance of heating and cooling that determines the thermal
phases of the H{\tt I}.  

The mean spin temperature should be interpreted not as a physical
temperature, but as a parameter that measures of the mixture of
WNM and CNM.  This is because the kinetic temperature in
the H{\tt I} varies over a very wide range,
from 20 K to 10$^4$ K \citep{Kulkarni_Heiles_1988}.  Blending
of warm and cool gas at the same velocity along the line of
sight to a background source gives a density-weighted harmonic
mean of this quantity,

\begin{equation}
\left< T_{sp} \right> \ = \ \frac{T_{EM}}{\left(1-e^{-\tau}\right)} \ =
\ \frac{\int{n \ ds}}{\int{\frac{n}{T_{kin}} \ ds}} 
\end{equation}

\noindent
where the integrals in the third expression are taken along the
line(s) of sight, $s$, including only gas with radial velocity that places
it into a given velocity channel or range of channels.
Making the assumption that the CNM temperature is roughly
constant at some typical value $T_{cool}$, and that the WNM
temperature is high enough that it does not contribute to
the optical depth in the denominator of equation 7, gives

\begin{equation}
\left< T_{sp} \right> \ = \ T_{cool} \ \frac{n_{WNM} + n_{CNM}}{n_{CNM}} 
\end{equation}

\noindent
where the ratio of CNM density to total H{\tt I} (WNM plus CNM) density in this
two-phase approximation can be denoted
$f_c = \frac{n_{CNM}}{n_{WNM} + n_{CNM}} = 1 - f_w$
\citep{Dickey_etal_2000}.  Detailed fitting of the line profile
shapes for corresponding features in the emission and absorption
spectra can give a good estimate for the CNM temperatures,
$T_{cool}$, which typically vary between 20 K and 100 K
for distinct absorption lines with peak optical depths of 0.1 or higher.
The mean value for $T_{cool}$ is about 50 K, as measured
by different techniques in different surveys with different
telescopes \citep{Heiles_Troland_2003b,Dickey_etal_2003}.
Similar analysis by \citet{Strasser_2006} of the absorption 
lines in the CGPS suggests that $T_{cool}$ does not change
significantly or systematically with increasing $R_g$.
The constancy of $T_{sp}$ with $R_g$ thus shows that
$f_c$ is also roughly constant in the outer Galactic disk.
Since the bottom panel of figure 5 shows a value of $\sim$400 K from the SGPS
data from 12 to 25 kpc in $R_g$, this gives $f_c$ 
in the range 10\% to 25\% for $T_{cool}$=40 K to 100 K.

This invariance of $f_c$ with $R_g$ is surprising in comparison
with the apparently strong variation of $f_c$ with $z$ in the
inner Galaxy.  The CNM and WNM seem to have quite different
distributions in $z$, with the WNM density showing a long tail
reaching to $|z| \sim 1$ kpc \citep{Lockman_Gehman_1991},
while the CNM has scale height of only 100 to 150 pc (\citealt{Crovisier_1978},
\citealt{Malhotra_1995}, but see \citealt{Stanimirovic_etal_2006} and
\citealt{Pidopryhora_Lockman_Shields_2007}
for studies of H{\tt I} clouds, at heights of 0.5 to 1 kpc in $z$,
that contain significant amounts of CNM).
The difference in these scale heights has been explained 
by variation in the gas pressure with $z$, which shifts the
equilibrium values of $n$ and $T_{cool}$ for which the
total heating and cooling rates are equal \citep{Wolfire_etal_2003}. 

One way to understand the difference between CNM scale heights in the inner and
outer Galaxy
is to identify a part of the CNM in the inner Galaxy with the molecular
cloud population.  The molecular clouds are the dominant form of ISM
mass in the inner Milky Way, particularly in the molecular ring
$3 < R_g < 6$ kpc.  In the outer disk the molecular phase
makes only a small contribution to the total ISM mass, with the
neutral and ionized phases playing the dominant role.  The molecular
clouds certainly have a different distribution from the atomic gas,
with the scale height of the disk being much narrower as traced in
CO than in H{\tt I}, for example.  Molecular gas is usually surrounded by
and/or mixed with partially atomic gas which shows 21-cm absorption
and so appears as CNM \citep{Goldsmith_Li_Krco_2007}.   Molecular
line cooling is so efficient that the H{\tt I} gas in a molecular cloud
is cooler than in the atomic-only CNM ($\sim$20 K vs. $\sim$50 K, \citealt{Goldsmith_2001}),
thus molecular clouds contribute disproportionately (compared
to their H{\tt I} column density) to 21-cm absorption
surveys of the inner Galaxy \citep[][section 5]{Kalberla_Kerp_2009}.  In
the outer Galaxy this source
of 21-cm absorption is insignificant, so
$f_c$ shows the mixing-ratio
of CNM to WNM without confusion by H{\tt I} in molecular clouds.
In the outer disk the evidence on figures 7a-7c suggests that
the scale heights of the WNM and CNM are nearly the same.

A more profound question is how the CNM can coexist with
the WNM as a familiar two-phase medium, in an environment
where the gas pressure, set by the depth of the disk
potential and the overburden of gas at higher $z$, is as
low as it must be in the far outer Galaxy.  We see from figure
5 that the density at $R_g \sim$20 to 25 kpc is a factor of
nearly 100 smaller than at the solar circle, and since the
temperatures are similar this means that the gas pressure is
less than 100 cm$^{-3}$ K.  The lowest value of the gas
pressure that allows the cool phase to exist, even in the
environment of the outer disk studied by \citet{Wolfire_etal_2003}, 
is $\sim$250 to 300 cm$^{-3}$ K.  Supplementing the pressure
of the gas with other forms such as magnetic pressure
does not necessarily help, since it is the collision rate
that sets the cooling in the equilibrium calculation, and
so heating-cooling equilibrium is set by product of the density and
random velocity, not by the total pressure.  

A similar theoretical problem is posed by cool H{\tt I} clouds in
the Magellanic Bridge and in the far outskirts of the Magellanic
Clouds \citep{Kobulnicky_Dickey_1999}, and
in many examples of HI absorption beyond
the edges of the disks of other galaxies.
Probably the explanation in most cases is that the
local space density, $n_H$, is increased by an order of
magnitude or more above the average, $\left< n_H \right>$,
due to shocks, either from supernova explosions or
from converging large scale gas flows ultimately driven 
by gravity.  The supernova rate is undoubtedly very low
in this environment, but 
converging flows may be common.
As long as the cooling time of the H{\tt I} is shorter
than the dynamical time scale of the large structures on the edges
of the Milky Way and other galaxies, 
the H{\tt I} may be driven from warm to cool and back to warm
as the pressure responds to the larger scale gas dynamics.

As \citet{Strasser_etal_2007} show, the spatial distribution
of the gas causing the absorption in the outer Galaxy is not
random; the CNM is gathered in large, coherent structures that
in turn follow patterns on the {\it l}-v diagram that may
connect to the spiral arms of the inner Galaxy.  This suggests
that departures from circular rotation caused by gravitational
perturbations initiate the runaway cooling that precipitates
the cool, atomic gas out of the WNM.

A powerful probe of the gas in and around galaxies
at high and intermediate redshifts is 21-cm absorption
\citep[reviewed by][]{Carilli_2006}.  An important question
in interpreting such lines is how to translate the 
measured optical depth to H{\tt I} column density.  The results
of this study suggest that using a value for $T_{sp}$ of 
250 to 400 K would be appropriate for the Milky Way seen
at an impact parameter of 10 to 25 kpc.  This is significantly
higher than the values of 100 to 150 K often used on the
basis of the relative abundances of CNM and WNM inside
the solar circle. 

\acknowledgments
We are grateful to F. J. Lockman, P. M. W. Kalberla, and an anonymous referee
for helpful suggestions.
This research was supported in part by NSF grant AST-0307603
to the University of Minnesota.
The Dominion Radio Astrophysical Observatory is operated as a
national facility by the National Research Council of Canada.
The Canadian Galactic Plane Survey is supported by a grant from
the National Sciences and Engineering Research Council of Canada.
The National Radio Astronomy Observatory is a facility of the 
National Science Foundation operated under cooperative agreement
by Associated Universities, Inc.
The ATCA and the Parkes Radio Telescope are part of the Australia Telescope,
which is funded by the Commonwealth of Australia for operation as a
National Facility managed by CSIRO.


\begin{deluxetable}{llcccccc}
\tablecaption{Galactic Plane Survey Parameters}
\tablecolumns{8}
\tablehead{

\colhead{Survey}&\colhead{area}&\colhead{angular}&\colhead{velocity}&
\colhead{rms noise}&\multicolumn{3}{c}{Number of Absorption}\\

&&\colhead{resolution}&\colhead{resolution}&\colhead{in $T_{EM}$}&
\multicolumn{3}{c}{Spectra with $\sigma_{\tau}$}\\

&&&&&\colhead{$<$0.02}&\colhead{0.02-0.05}&\colhead{0.05-0.1}\\ 
}
\startdata

VGPS\tablenotemark{a}
& 18\arcdeg$ < l < 67$\arcdeg&1\arcmin &1.56 km/s &2K&15&64&49\\
& -1.3\arcdeg$ < b < $+1.3\arcdeg&&&&&&\\

CGPS\tablenotemark{a}
& 65\arcdeg$ < l < $175\arcdeg&1\arcmin &1.32 km/s&3 K&32&108&256\\
& -3.6\arcdeg$ < b < $+5.6\arcdeg&&&&&&\\

SGPS\tablenotemark{b}
&253\arcdeg$ < l < 358$\arcdeg&2\arcmin &1.0 km/s &1.6 K&30&42&54\\
& -1.5\arcdeg$ < b < $+1.5\arcdeg&&&&&&\\

\enddata
\tablenotetext{a}{Data are available from
http://www2.cadc-ccda.hia-iha.nrc-cnrc.gc.ca/cgps/
}
\tablenotetext{b}{Data are available from
http://www.atnf.csiro.au/research/HI/sgps/queryForm.html
}

\end{deluxetable}


\begin{figure}

\includegraphics[angle=-90,width=8cm]{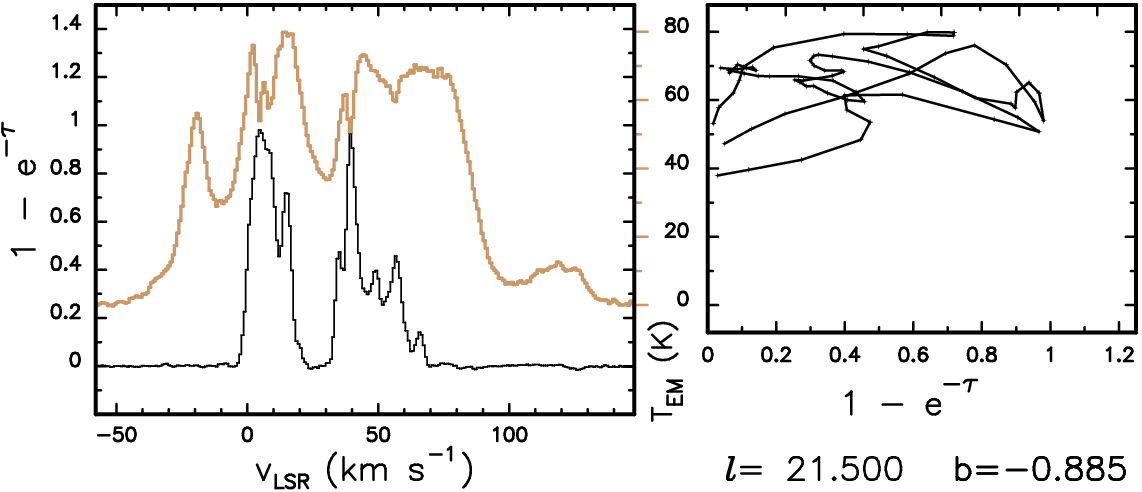}
\includegraphics[angle=-90,width=8cm]{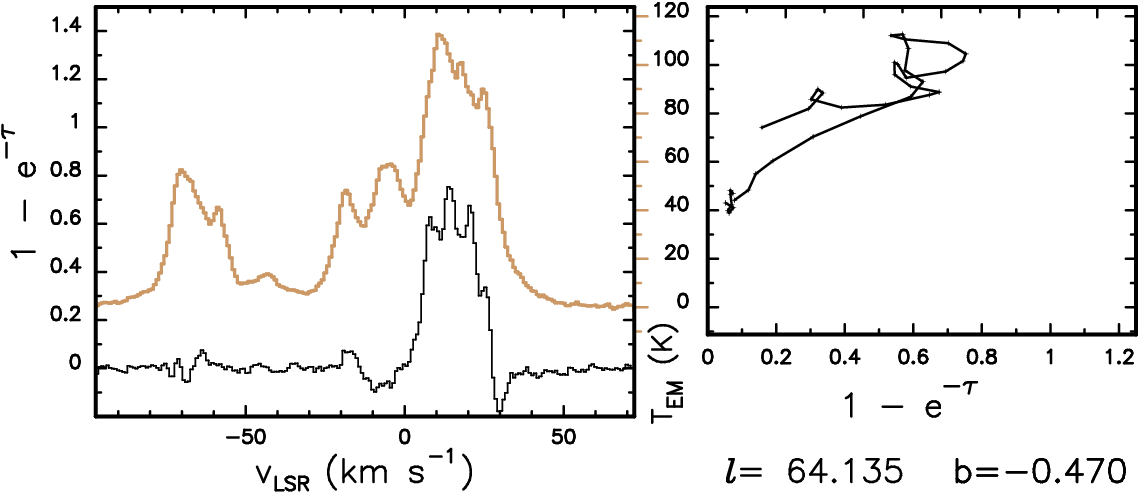}

\includegraphics[angle=-90,width=8cm]{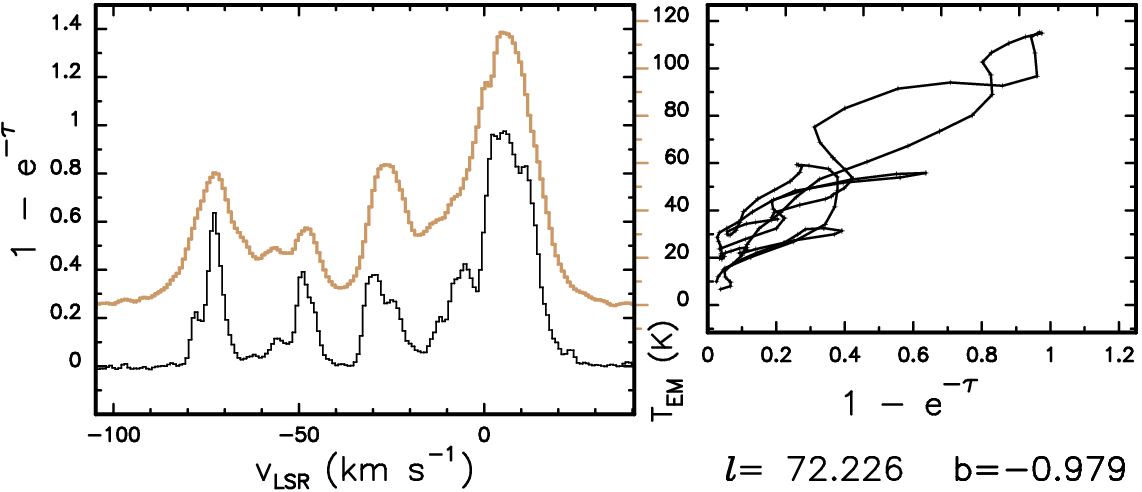}
\includegraphics[angle=-90,width=8cm]{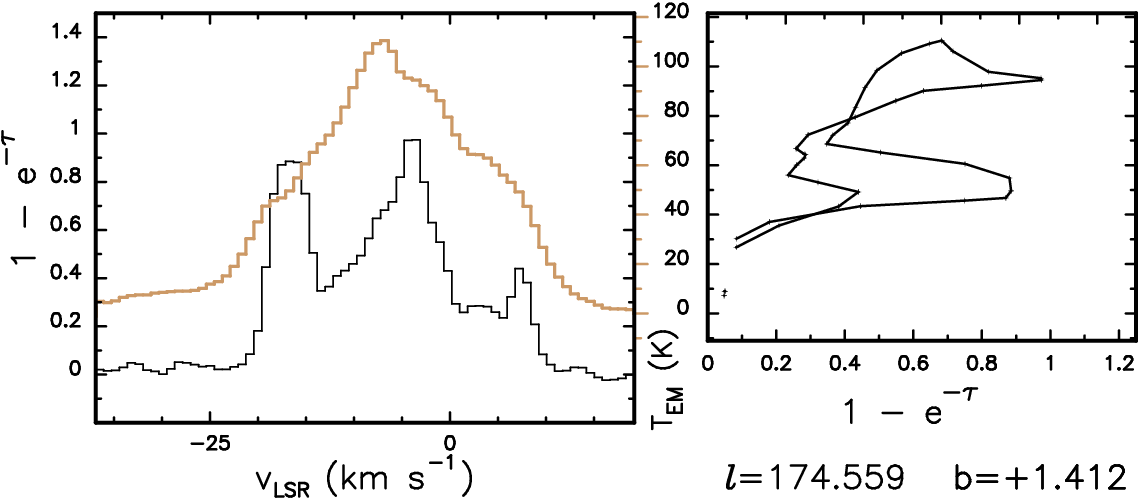}

\includegraphics[angle=-90,width=8cm]{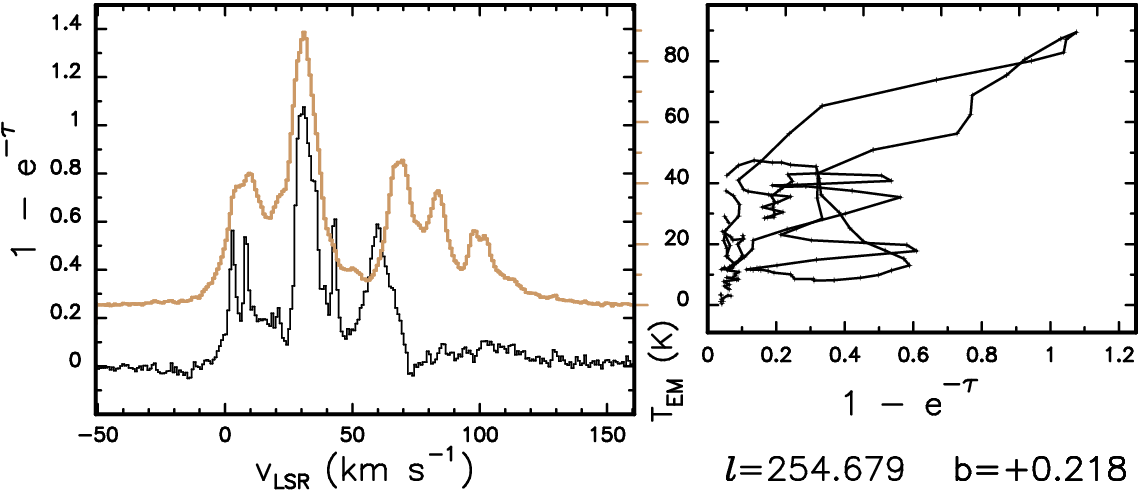}
\includegraphics[angle=-90,width=8cm]{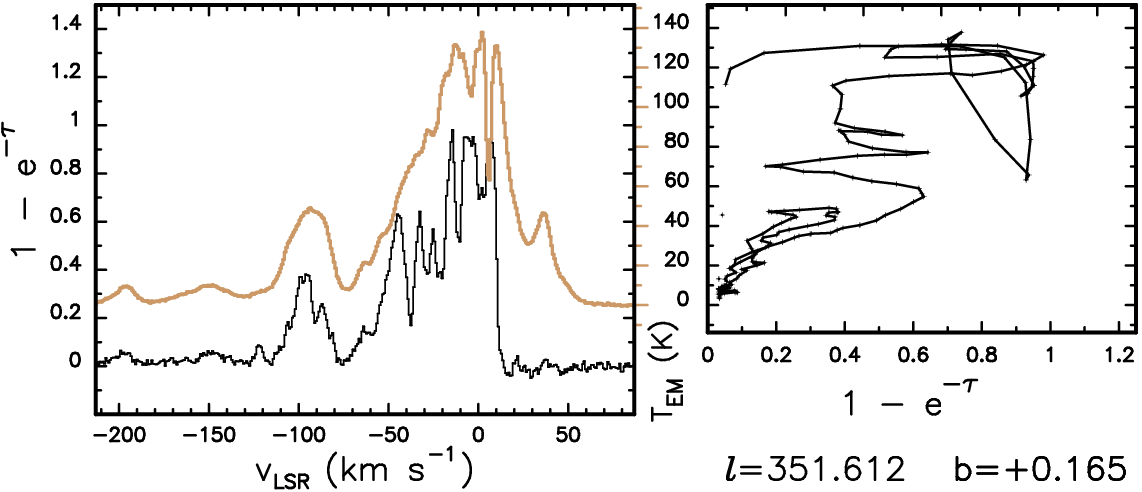}

\caption{Sample spectra from the three surveys with $\sigma_{\tau} < 0.02$.
Top row shows the first and last spectrum in this group from the VGPS,
the middle row shows the first and last spectrum in this group from
the CGPS, and the bottom row shows the first and last spectrum in
this group from the SGPS.  The left panel shows the absorption spectrum
plotted as $\left( 1 - e^{-\tau} \right)$, with the corresponding
vertical scale on the left, and the emission spectrum, $T_{EM}$, offset
up and scaled on the right hand axis, between the two panels.
In the electronic edition the emission is shown in gold.
The right hand panel shows the emission plotted vs. the absorption,
with the y axis again the scale for $T_{EM}$, and the x axis
the scale for $\left( 1 - e^{-\tau} \right)$.
In the electronic edition all spectra are included, from
all three surveys.  In the electronic edition 
figure 1 includes 15 pairs of spectra
from the VGPS, 32 from the CGPS, and 30 from the SGPS.
}
\end{figure}

\begin{figure}

\includegraphics[angle=-90,width=8cm]{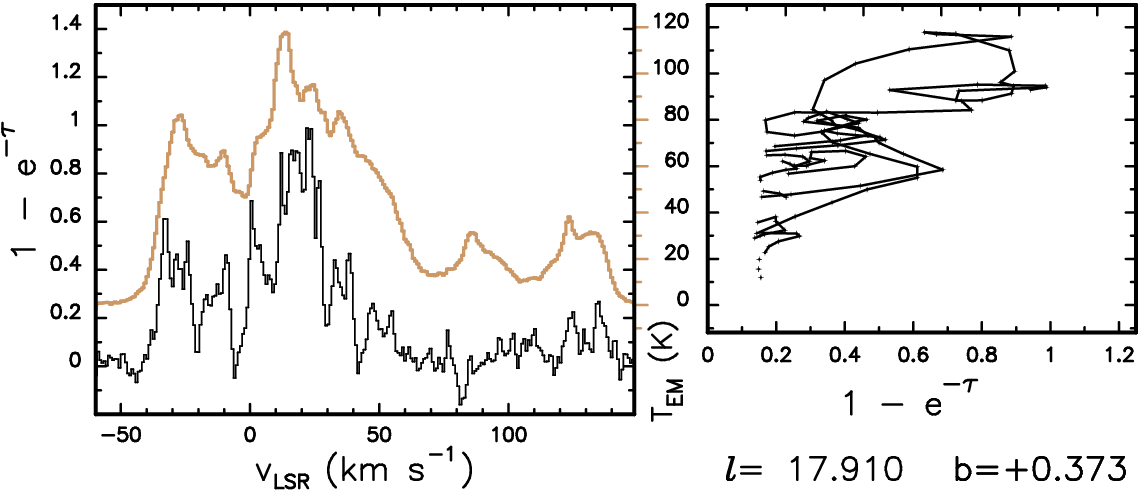}
\includegraphics[angle=-90,width=8cm]{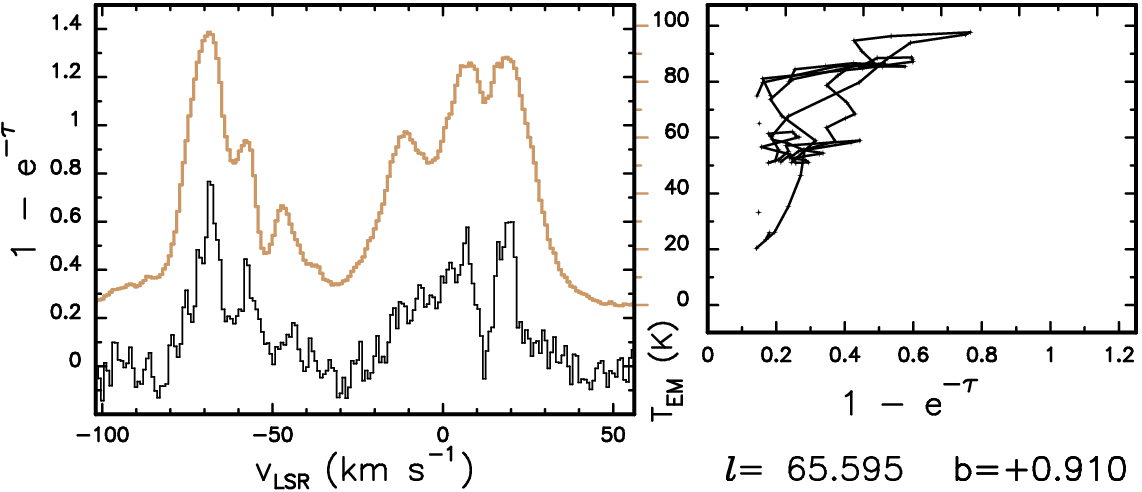}

\includegraphics[angle=-90,width=8cm]{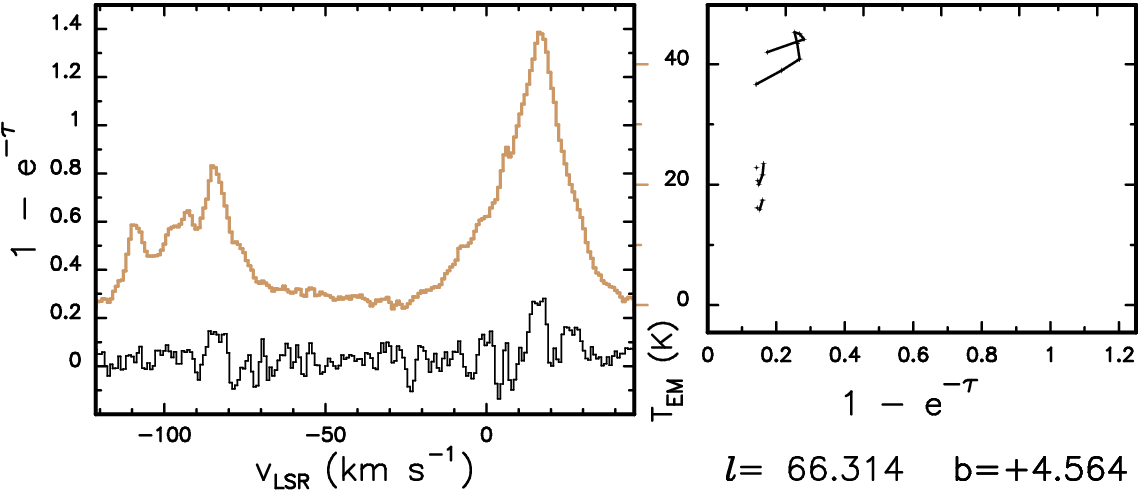}
\includegraphics[angle=-90,width=8cm]{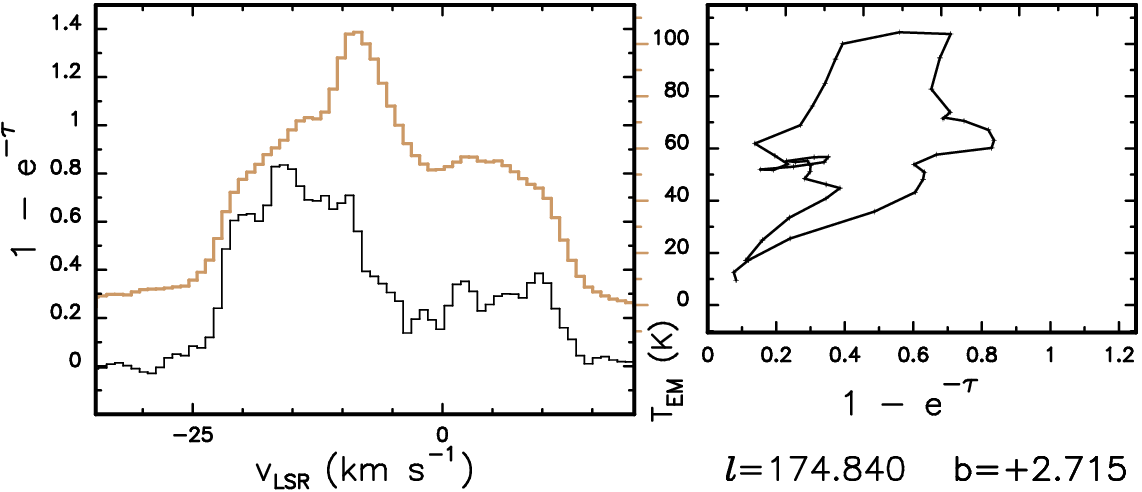}

\includegraphics[angle=-90,width=8cm]{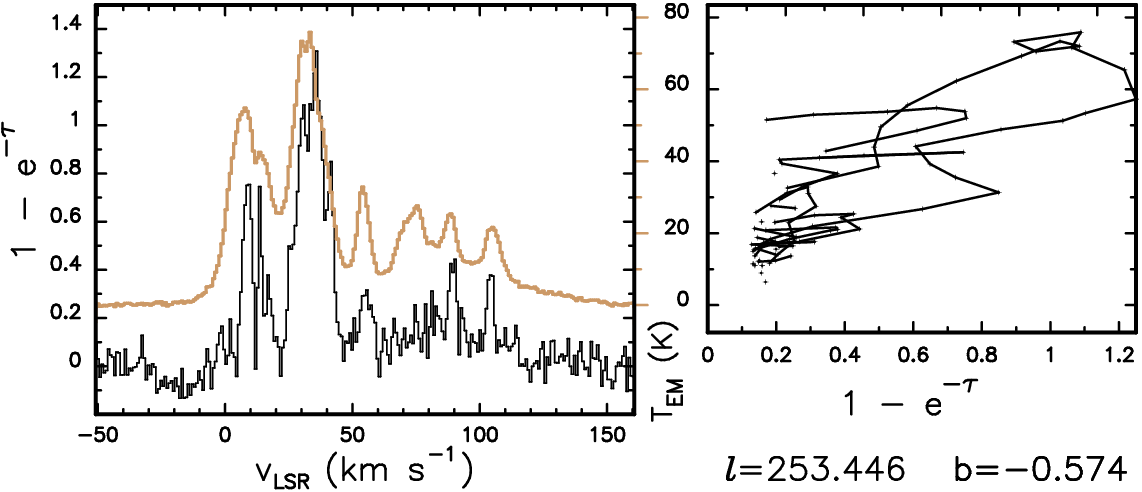}
\includegraphics[angle=-90,width=8cm]{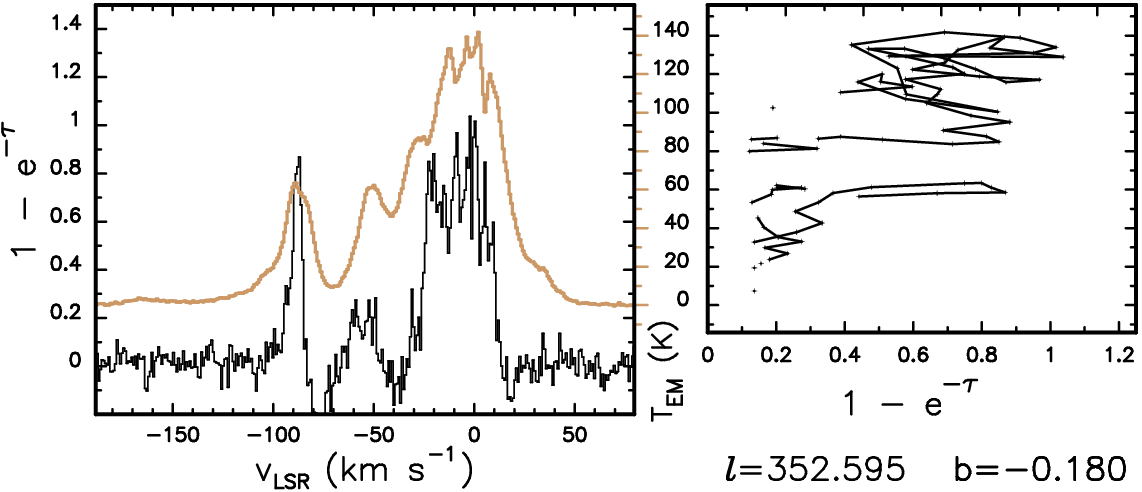}

\caption{Sample spectra from the three surveys with $0.02 \leq \sigma_{\tau}
< 0.05$.
The layouts of the panels are the same as in figure 1, with the first
and last spectra from the VGPS on the top line, the first and last from
the CGPS on the second line, and the first and last of the SGPS on the
third line.   In the electronic edition figure 2 includes 64 pairs of spectra
from the VGPS, 108 from the CGPS, and 42 from the SGPS.
}

\end{figure}

\begin{figure}

\includegraphics[angle=-90,width=8cm]{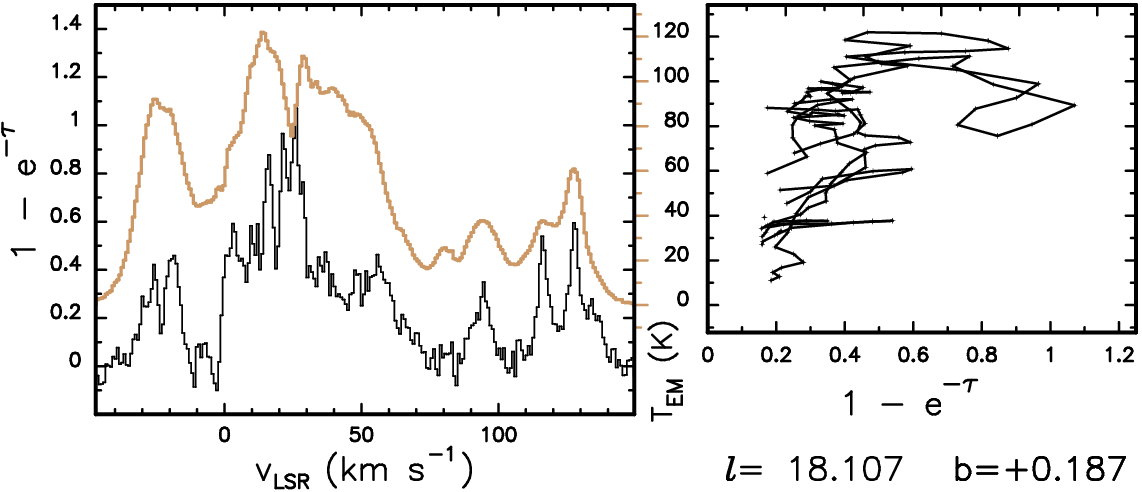}
\includegraphics[angle=-90,width=8cm]{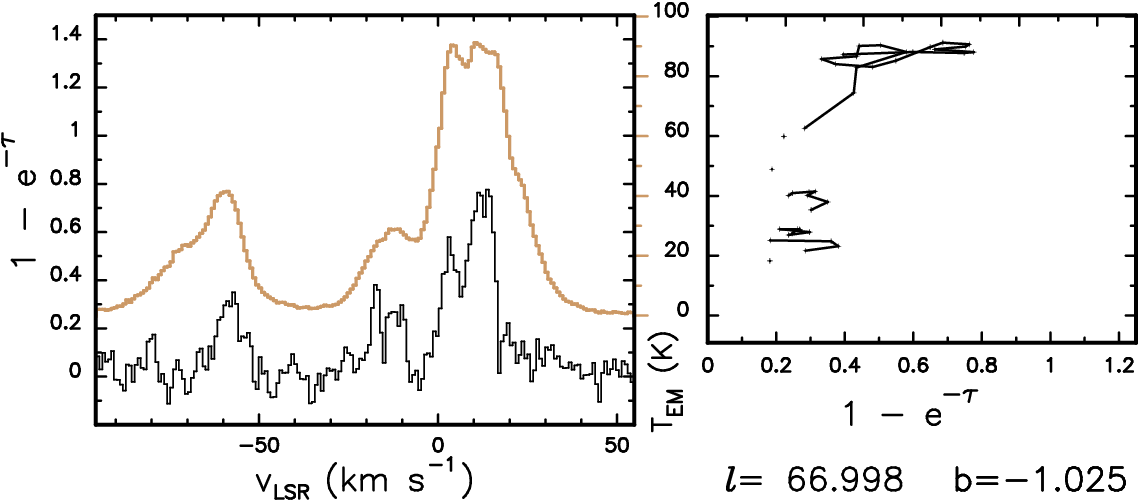}

\includegraphics[angle=-90,width=8cm]{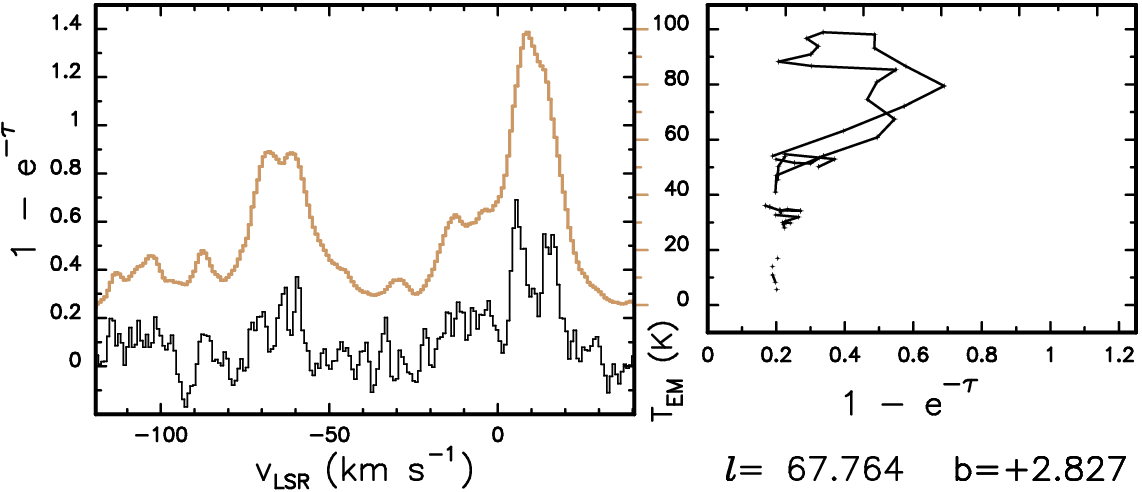}
\includegraphics[angle=-90,width=8cm]{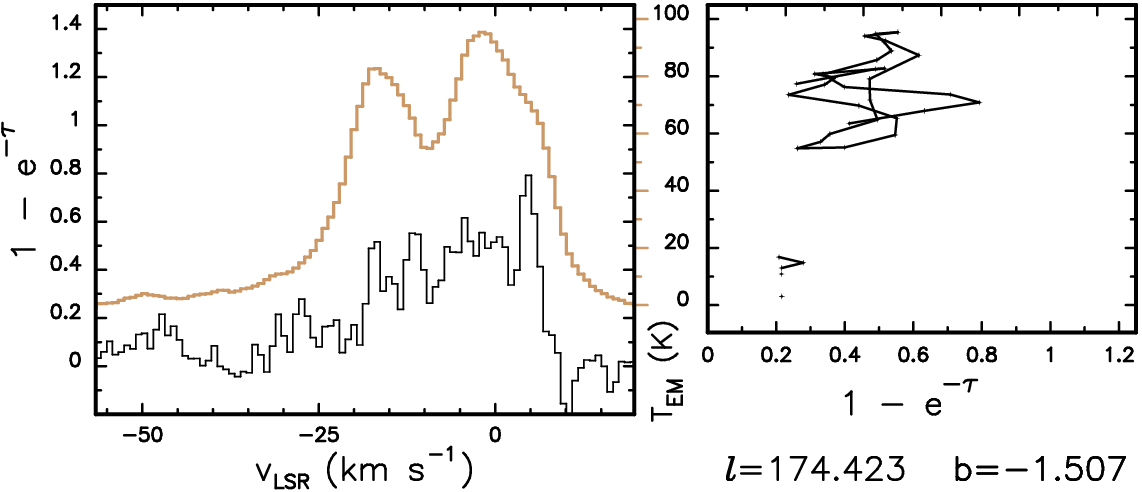}

\includegraphics[angle=-90,width=8cm]{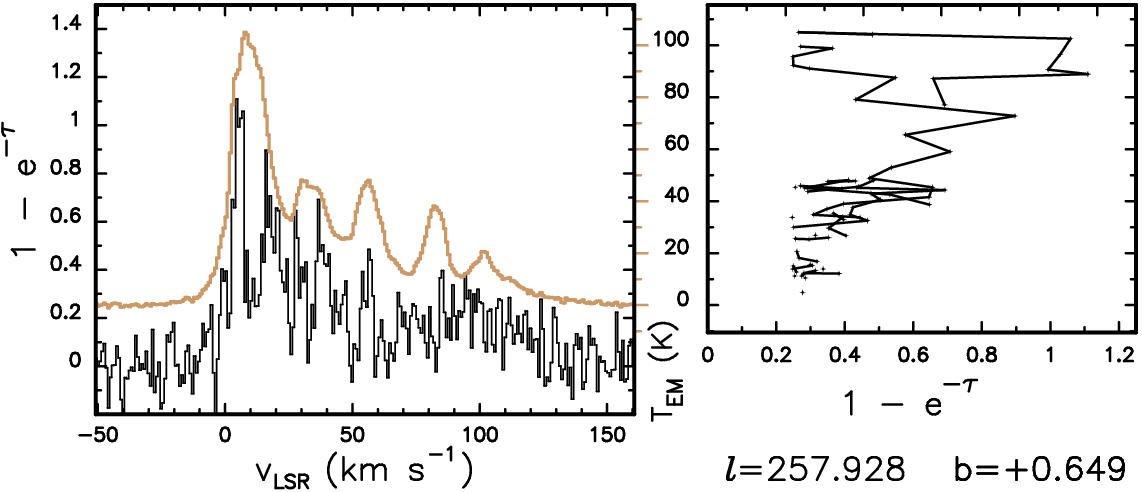}
\includegraphics[angle=-90,width=8cm]{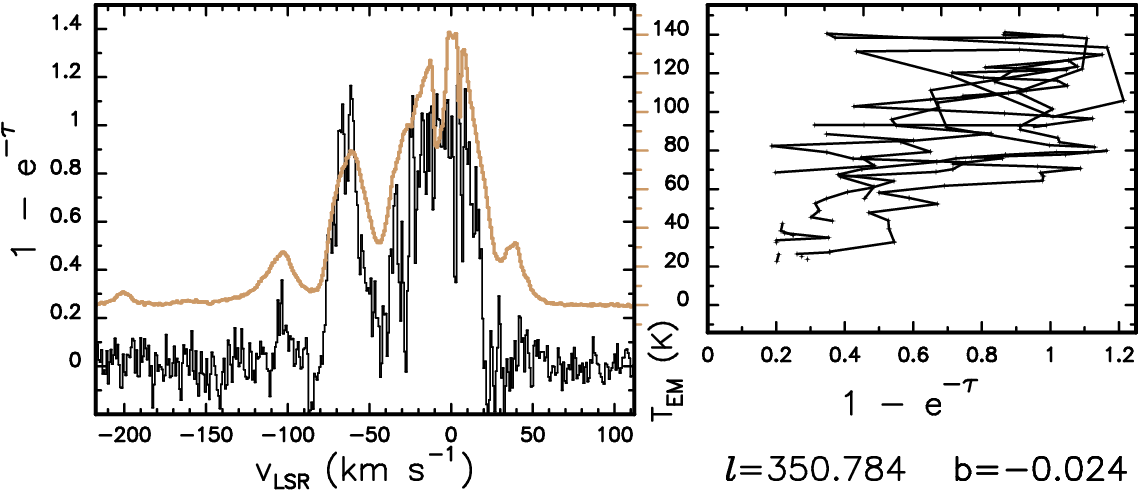}

\caption{Sample spectra from the three surveys with $0.05 \leq \sigma_{\tau}
< 0.10$.
The layouts of the panels are the same as in figure 1, with the first
and last spectra from the VGPS on the top line, the first and last from
the CGPS on the second line, and the first and last of the SGPS on the
third line.  In the electronic edition figure 3 includes
64 pairs of spectra from the VGPS, 108 from the CGPS, and
42 from the SGPS. }
\end{figure}

\begin{figure}



\includegraphics[angle=-90,width=8cm]{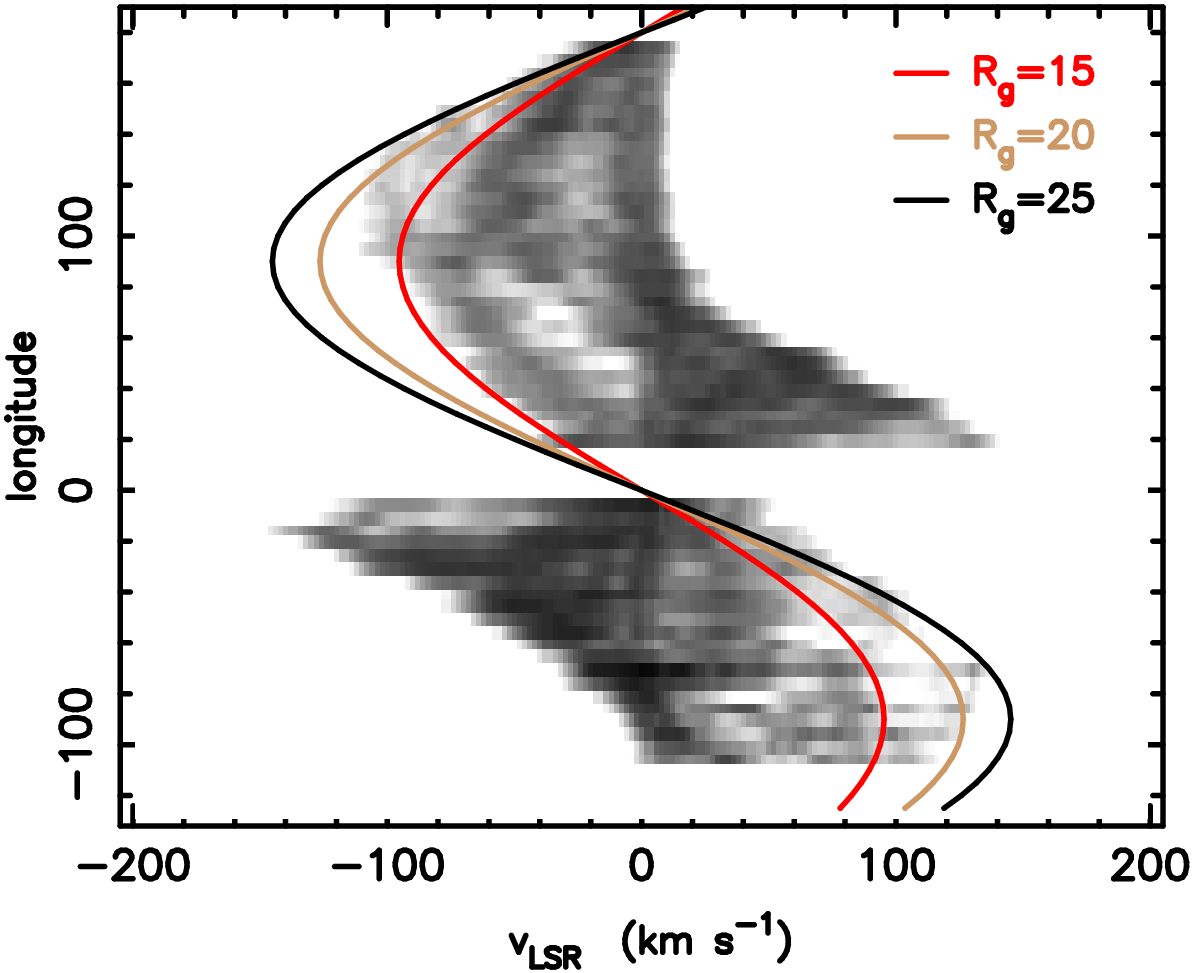}
\includegraphics[angle=-90,width=8cm]{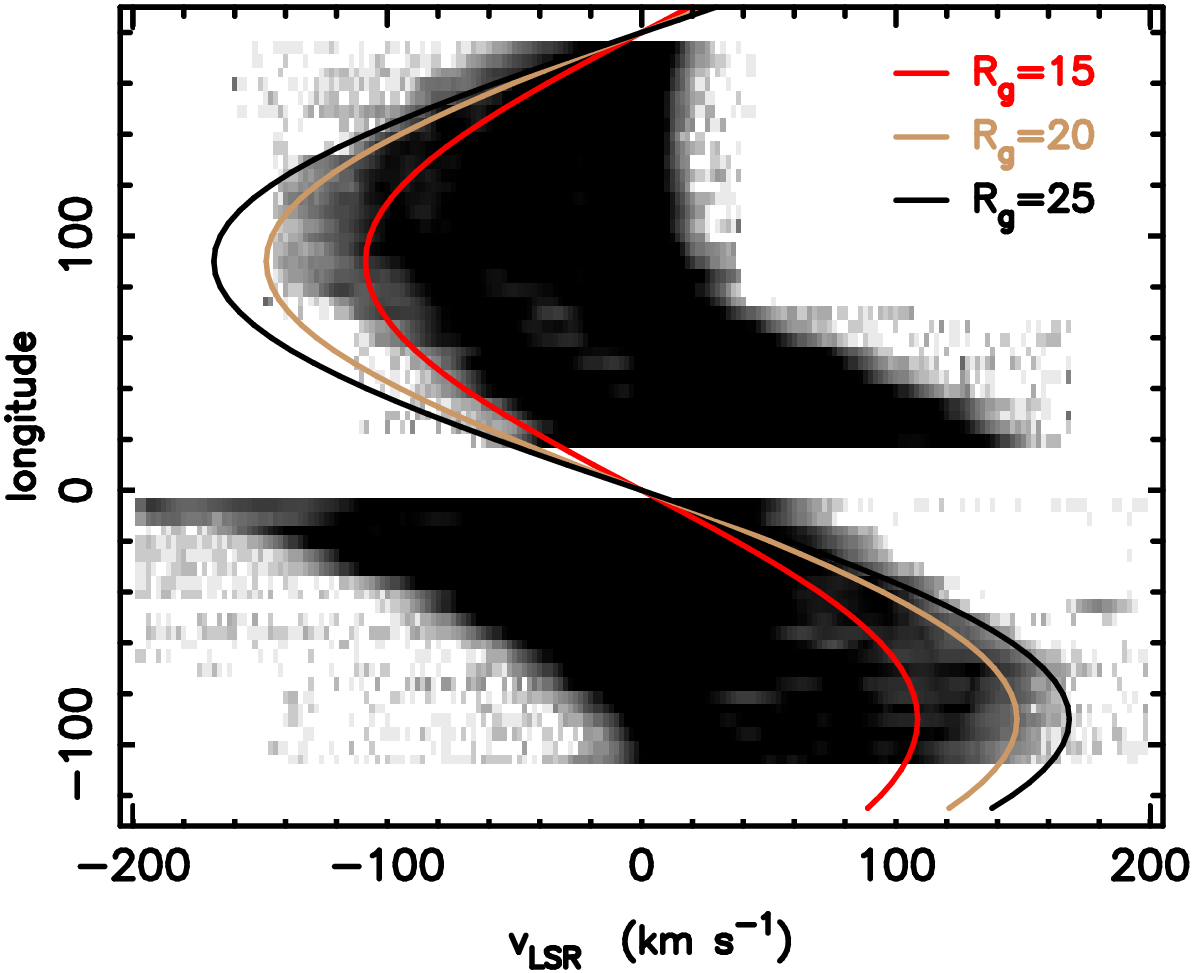}

\includegraphics[angle=-90,width=8cm]{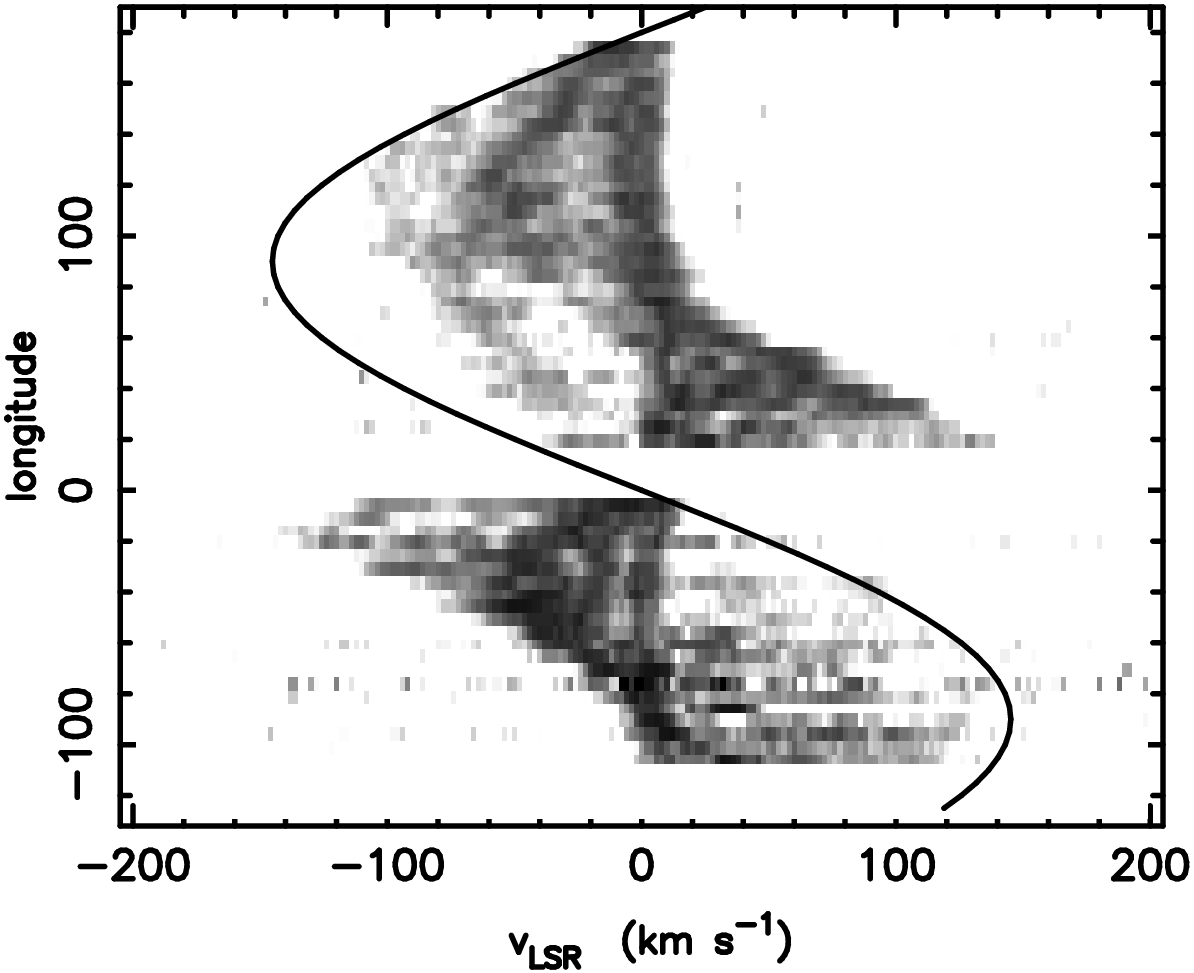}
\includegraphics[angle=-90,width=8cm]{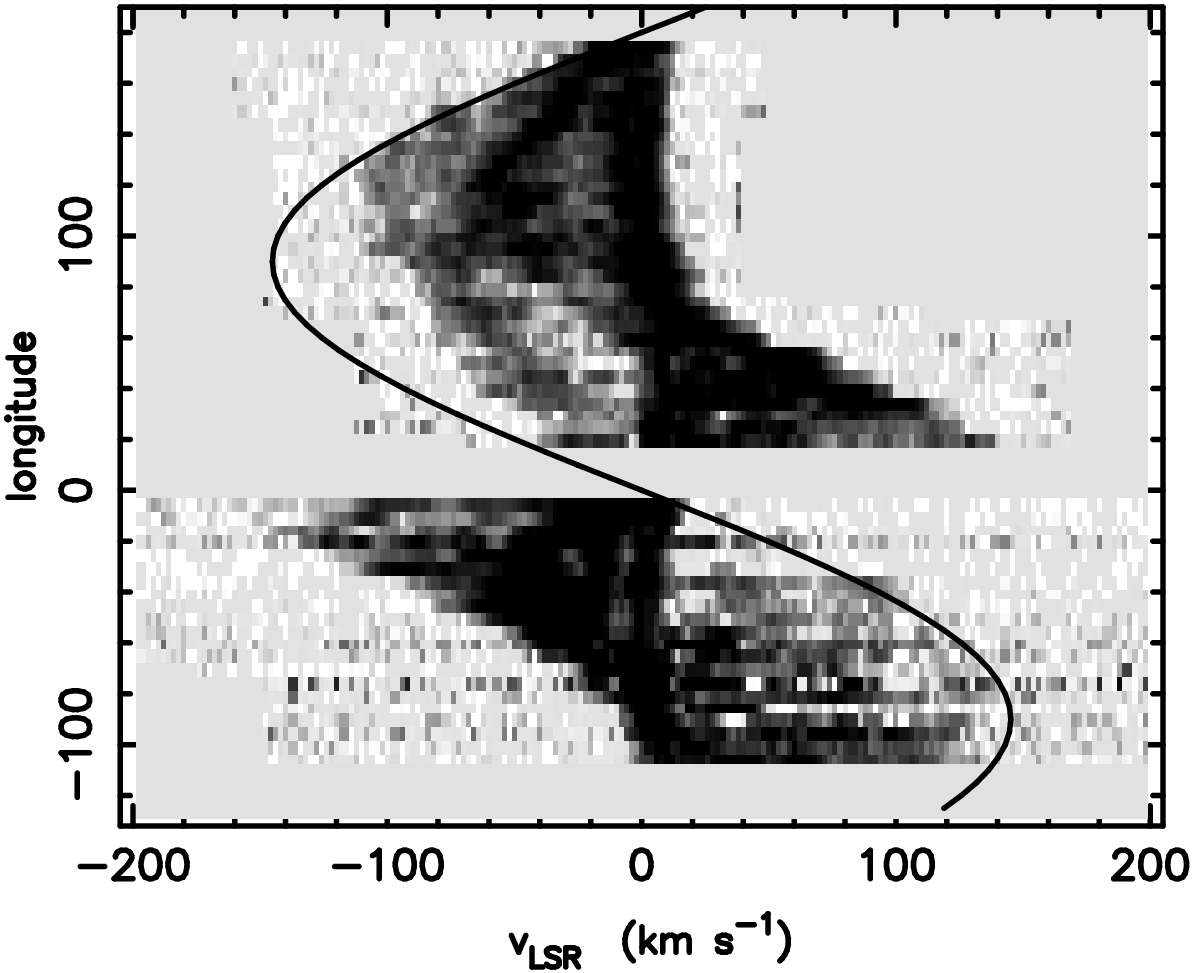}

\caption{Longitude-velocity diagrams for emission, $T_{EM}$, and absorption,
$\left( 1 - e^{-\tau} \right)$.  The upper panels show the emission, with
two different levels of grey scale.  The lower panels show the absorption.
The upper panels have lines of constant $R_g$ indicated.  The left panel
has the prediction of the flat rotation curve used here (equations 1 - 3),
the right panel shows similar curves for a rotation curve that stays 
flat to 15 kpc and then decreases as a Keplerian.  The left hand
panels have a shallower grey-scale transfer function than the right hand panels,
to better show the structure at different levels.  The lower panels
have the curve for $R_g$=25 kpc from the flat rotation curve
repeated from the upper left panel.
}
\end{figure}

\begin{figure}
\hspace{3cm}
\includegraphics[angle=-90,width=10cm]{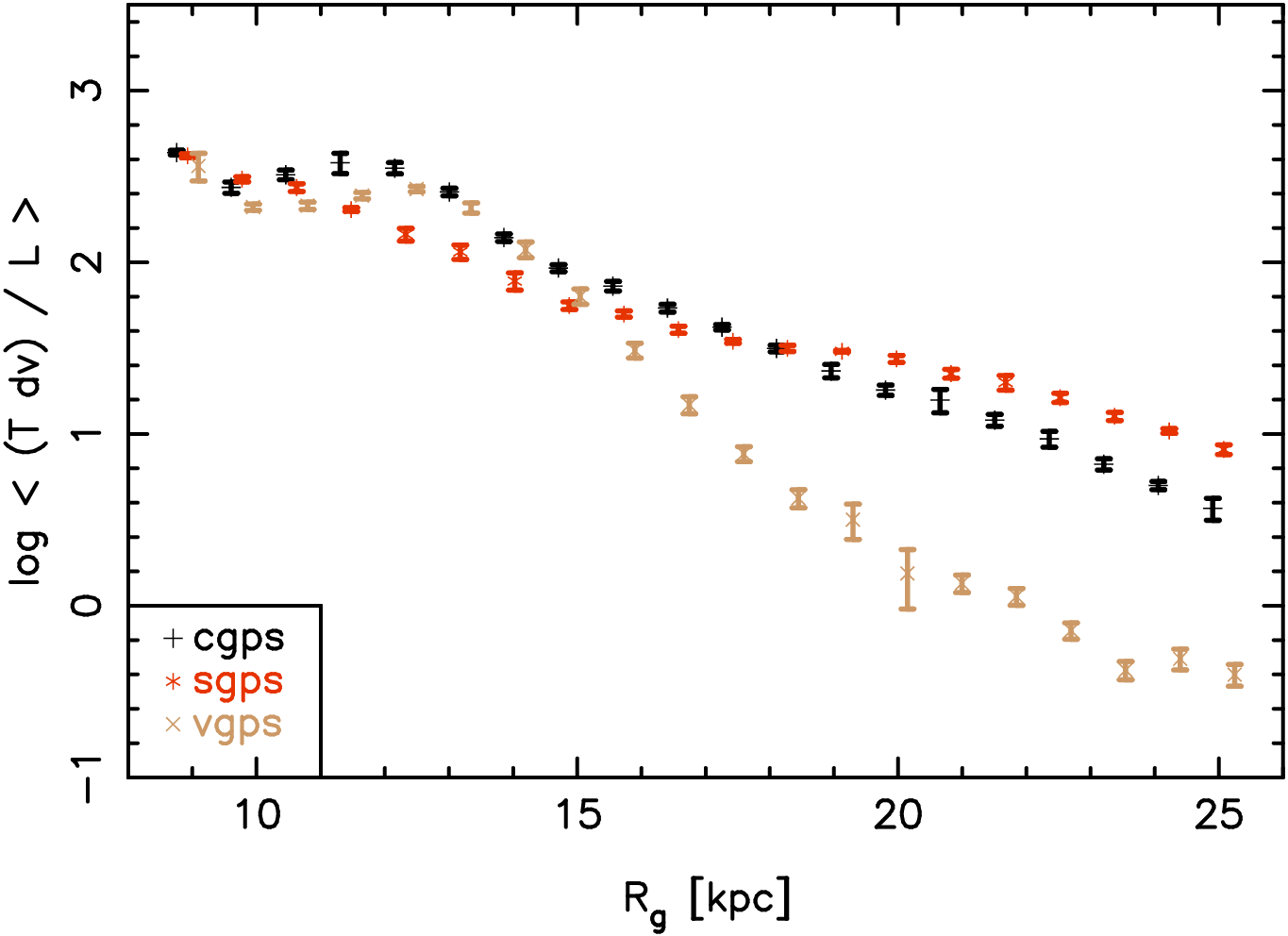}

\hspace{3cm}
\includegraphics[angle=-90,width=10cm]{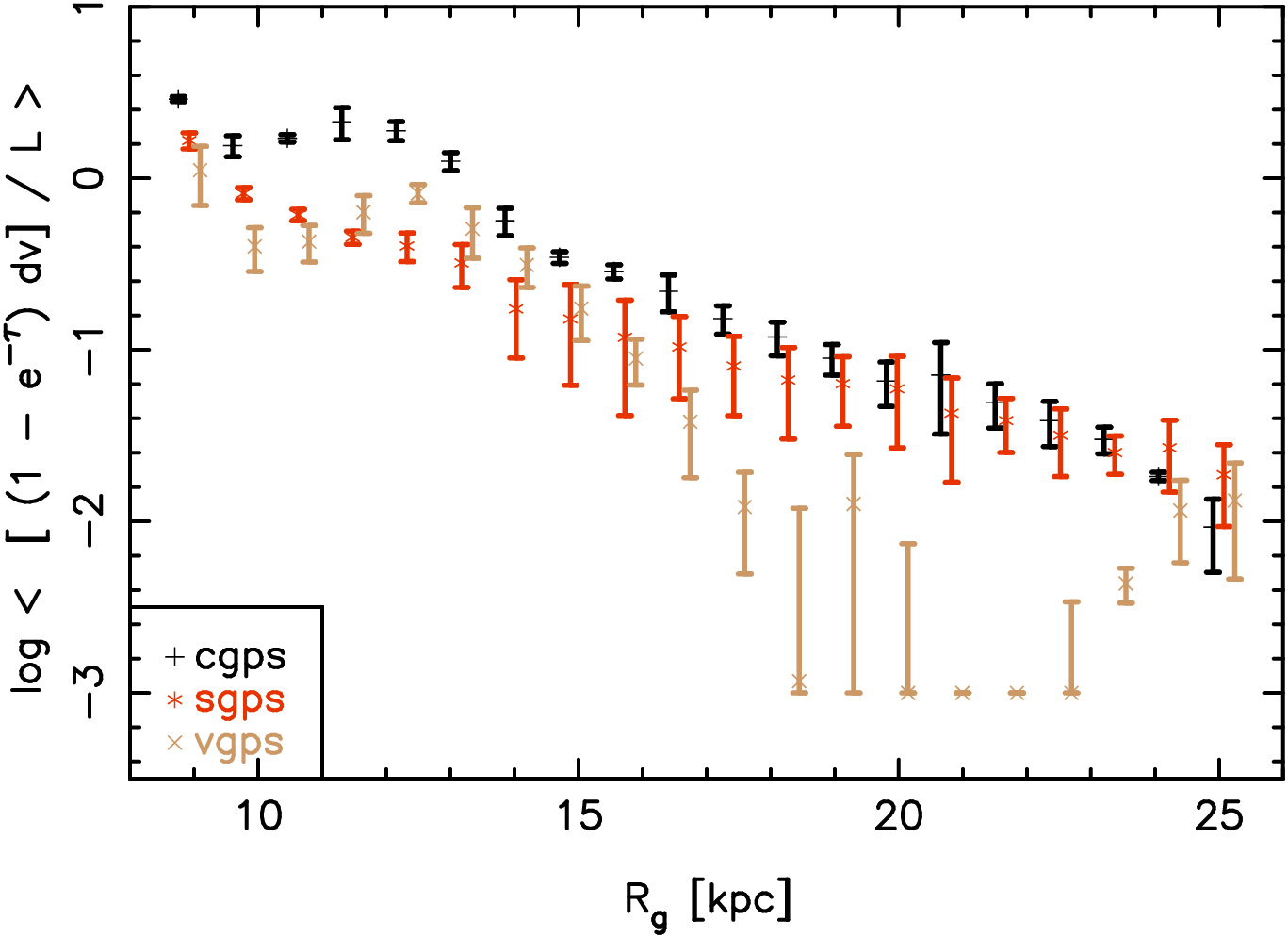}

\hspace{3cm}
\includegraphics[angle=-90,width=10cm]{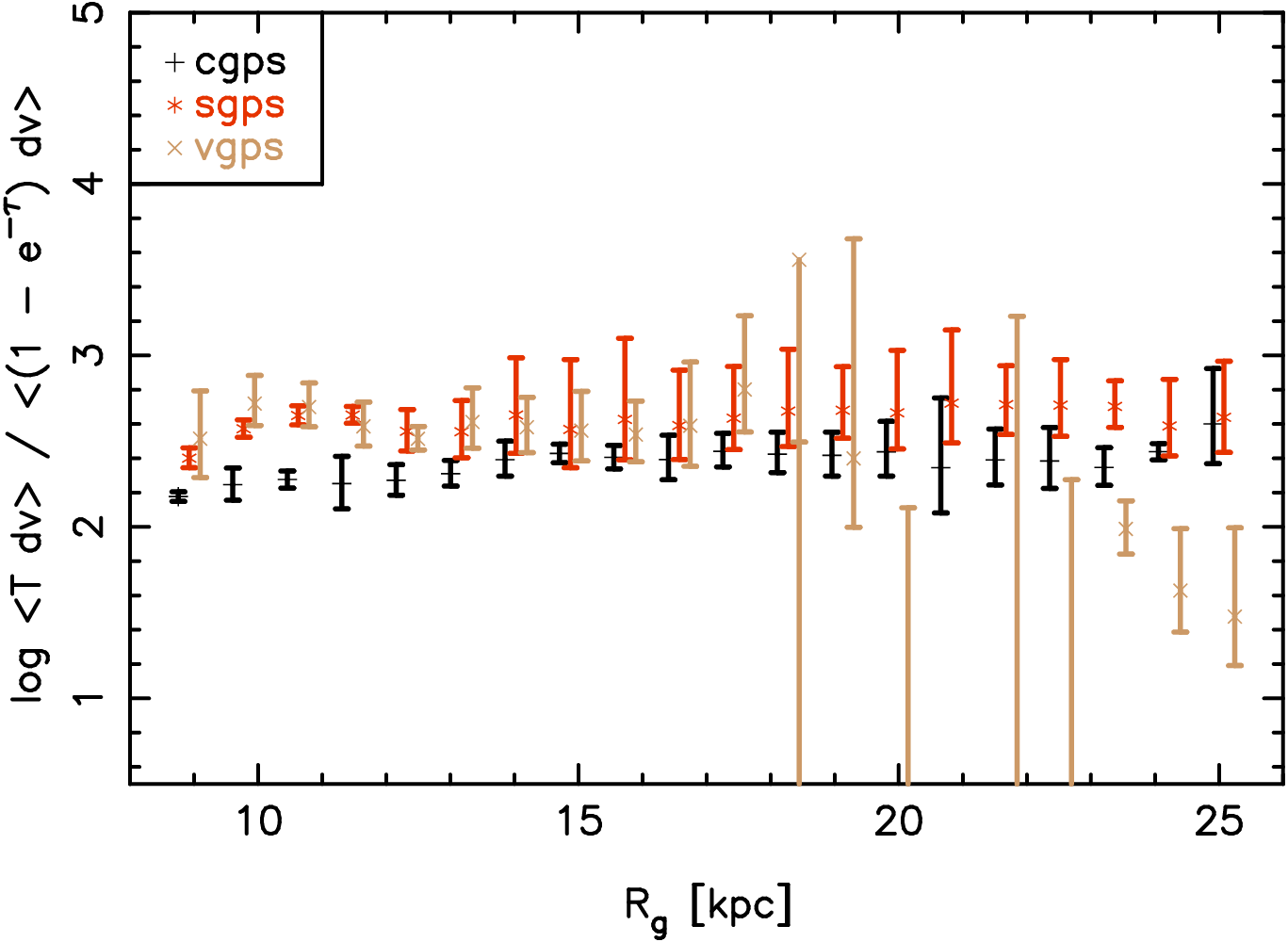}

\caption{The radial dependence of the 21-cm emission per
unit path length (top panel, proportional to $\left< n_H \right>$),
the 21-cm absorption per
unit path length ($\left< \kappa \right>$, middle panel),
and the ratio of emission to absorption, ($T_{sp}$, bottom panel).
}
\end{figure}

\begin{figure}
\hspace{3.5cm}
\includegraphics[angle=-90,width=8cm]{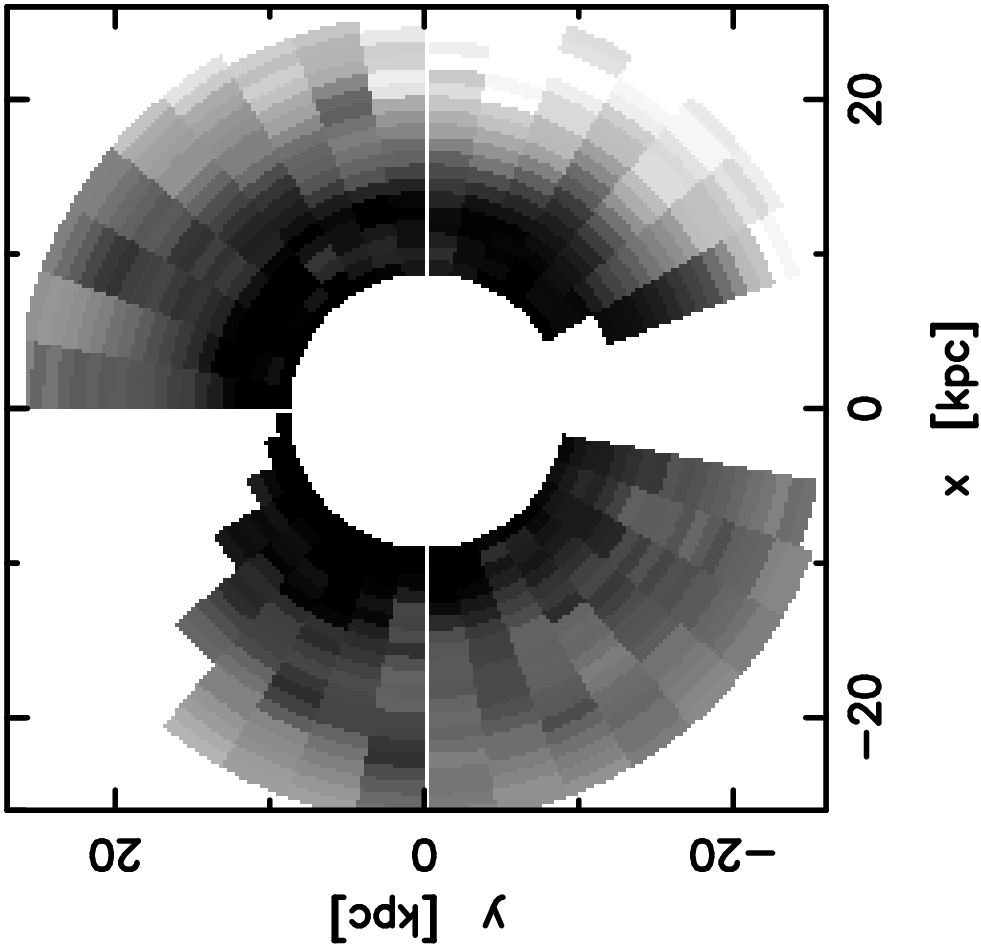}

\vspace{.5cm}

\hspace{3.5cm}
\includegraphics[angle=-90,width=8cm]{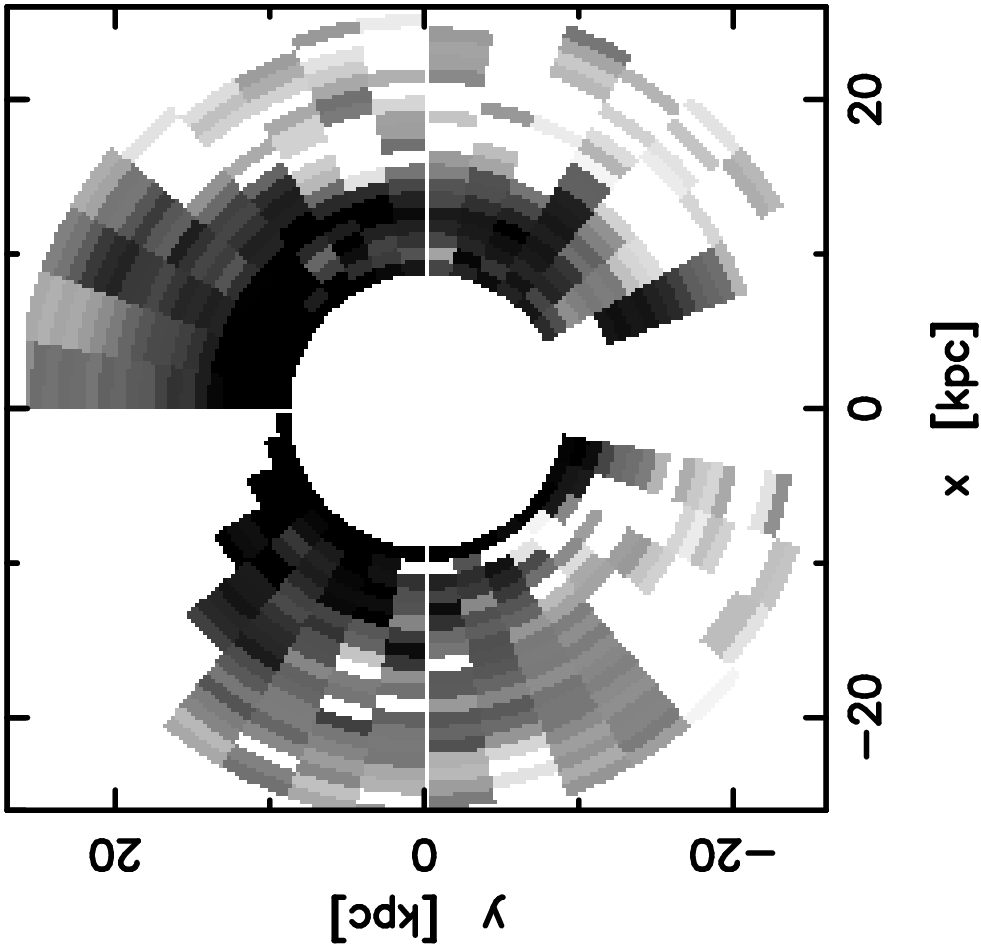}

\caption{The azimuthal distribution of the emission, shown 
on the upper panel (a), and the opacity shown
on the lower panel (b).  The effect of the warp is evident 
in the empty region in the lower right.  The Perseus Arm is clearly
visible in the first and second quadrants as the
dark feature just outside the solar circle.  The grey scale is
logarithmic, with range +2.5 (black) to -0.5 (white) 
in the upper panel, and 0 to -2.5 in the lower panel, with
units as in the upper two panels of figure 5.
The axes are labelled in kpc, assuming $R_o = 8.5$ kpc.
}
\end{figure}

\begin{figure}
\hspace{3.5cm}
\includegraphics[angle=-90,width=8cm]{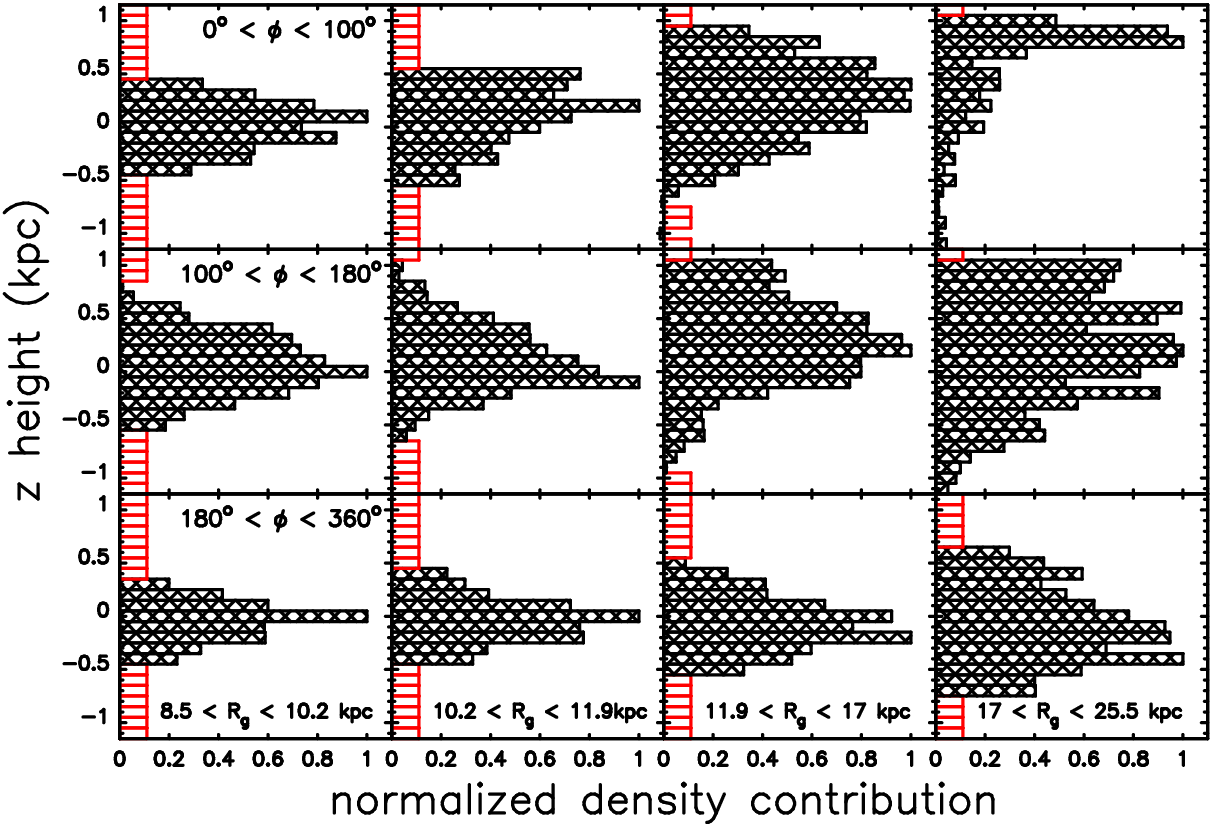}

\vspace{.5cm}

\hspace{3.5cm}
\includegraphics[angle=-90,width=8cm]{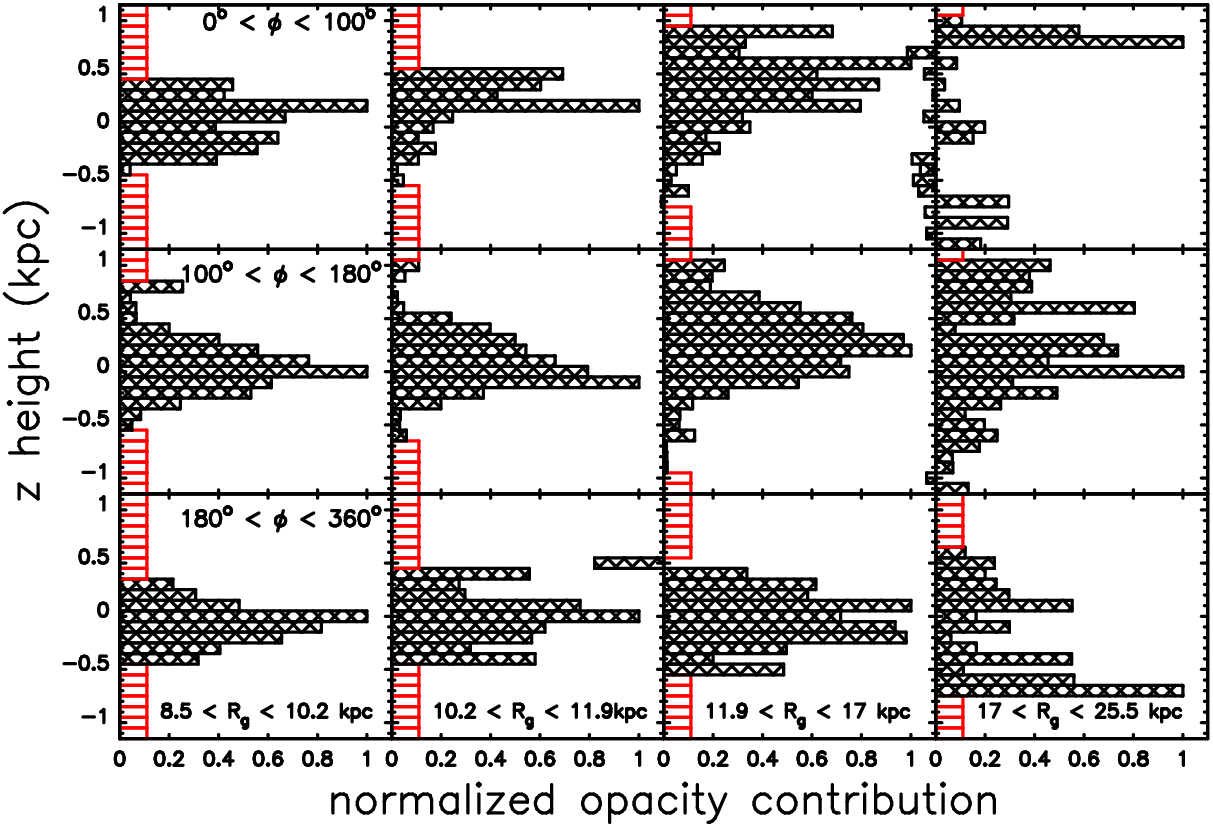}

\caption{Histograms of the distributions of HI density and opacity with $z$ in
different regions of the outer Milky Way disk.  The shaded rectangles are 
histograms, rotated so that the $z$ axis is vertical, where the height of
each bar represents the amount of gas in each 100 pc wide bin in z, from -1.1
to +1.1 kpc.  The unshaded rectangles (red in the electronic edition) mark 
regions where the number of samples is too small to give reliable results.
The different panels contain data from separate regions, with the rows showing
different ranges of azimuth, $\phi$, and the columns showing different 
ranges of Galactic radius, $R_g$, as indicated.  The upper figure (a)
shows the distribution of HI density, measured by the emission per unit
path length, and the lower panel (b) shows the opacity, $\left< \kappa \right>$.
The flaring of the disk in evident as a widening of the distributions with
increasing $R_g$ at all azimuths.  The warp is most clearly seen in the top
row, as the median $z$ increases with increasing $R_g$ in both the emission
and the absorption.  In all the panels the histograms have been scaled so
that the largest bar has unit height.  This means that the rightmost column
has scale factor more than 30 times larger than the leftmost column in absolute
units. 
}
\end{figure}

\end{document}